\documentclass[showpacs,twocolumn,aps,preprintnumbers,letterpaper]{revtex4}
\usepackage{amsmath,amssymb}
\usepackage{epsfig}
\usepackage{graphicx}
\usepackage{amsmath}
\usepackage{slashed}
\usepackage{amsfonts}
\usepackage{epstopdf}

\newcommand{\be}{\begin{equation}}
\newcommand{\ee}{\end{equation}}
\newcommand{\bear}{\begin{eqnarray}}
\newcommand{\eear}{\end{eqnarray}}
\newcommand{\ba}{\begin{array}}
\newcommand{\ea}{\end{array}}

\def\be{\begin{eqnarray}}
\def\ee{\end{eqnarray}}
\def\bea{\be}
\def\eea{\ee}

\newcommand{\e}{{\mbox{e}}}

\def\roughly#1{\mathrel{\raise.3ex\hbox{$#1$\kern-.75em%
\lower1ex\hbox{$\sim$}}}}

\begin{document}

\title{Light Adjoint Quarks  in the  Instanton-Dyon Liquid Model IV}

\author{Yizhuang Liu}
\email{yizhuang.liu@stonybrook.edu}
\affiliation{Department of Physics and Astronomy, Stony Brook University, Stony Brook, New York 11794-3800, USA}

\author{Edward Shuryak}
\email{edward.shuryak@stonybrook.edu}
\affiliation{Department of Physics and Astronomy, Stony Brook University, Stony Brook, New York 11794-3800, USA}

\author{Ismail Zahed}
\email{ismail.zahed@stonybrook.edu}
\affiliation{Department of Physics and Astronomy, Stony Brook University, Stony Brook, New York 11794-3800, USA}


\date{\today}
\begin{abstract}
We discuss  the instanton-dyon liquid model with $N_f$ Majorana quark flavors
in the adjoint representation of color $SU_c(2)$ at finite temperature.  We briefly recall the index theorem on $S^1\times R^3$
for twisted adjoint fermions in a BPS dyon background of arbitrary holonomy, and use the ADHM
construction to explicit the  adjoint anti-periodic zero modes.
We use these results to derive the partition function of an interacting instanton-dyon
ensemble with $N_f$ light and anti-periodic adjoint quarks. We develop the model in details  by mapping the theory on a 3-dimensional quantum effective theory with adjoint quarks with manifest $SU(N_f)\times Z_{4N_f}$ symmetry. Using a mean-field
analysis at weak coupling and strong screening, we show that  center symmetry requires  the spontaneous breaking of chiral symmetry,
which is shown to only take place for $N_f=1$.
For a sufficiently dense liquid, we find that the ground state is center symmetric and breaks spontaneously  flavor symmetry
through $SU(N_f)\times Z_{4N_f}\rightarrow O(N_f)$. As the liquid dilutes with increasing temperature,
center symmetry and chiral symmetry are restored. We present numerical and analytical estimates for the transition temperatures.
\end{abstract}
\pacs{11.15.Kc, 11.30.Rd, 12.38.Lg}


\maketitle

\setcounter{footnote}{0}


\section{Introduction}

This work is a continuation of our earlier studies~\cite{LIU1,LIU2,LIU3}  of the  gauge
topology in the confining phase of a theory with the simplest gauge group $SU(2)$.
We suggested that the confining phase below the transition temperature is an
   ``instanton dyon" (and anti-dyon) plasma which is dense enough
to generate strong screening. The dense plasma is amenable to
standard mean field methods.

The key ingredients in the instanton-dyon liquid model are the so called KVBLL
instantons threaded by finite holonomies~\cite{KVLL}  splitted into their constituents, the instanton-dyons.
Diakonov and Petrov~\cite{DP,DPX}  have shown that the KvBLL instantons dissociate in the
confined phase and recombine in the deconfined phase, using solely the BPS protected moduli space.
The inclusion of the non-BPS induced interactions, 
through the so called streamline set of configuration, is important numerically, but it does not alter this observation
\cite{LARSEN}. The dissociation of instantons into constituents was advocated originally by Zhitnitsky
and others~\cite{ARIEL}.  Unsal and collaborators~\cite{UNSAL} proposed a specially tuned setting in which
instanton constituents (they call instanton-monopoles) induced confinement even at exponentially
small densities, at which the semi-classical approximations is parametrically accurate. Key feature of this setting
is cancellation of the perturbative Gross-Pisarski-Yaffe holonomy potential.

The KvBLL instantons  fractionate into constituents with fractional topological charge $1/N_c$.
Their fermionic zero modes do not fractionate but rather migrate between various  constituents~\cite{KRAAN}.
This interplay between the zero modes and the constituents is captured precisely by the Nye-Singer
index theorem~\cite{SINGER}. For fundamental fermions, we have recently shown in the
mean-field approximation that the center
symmetry and chiral symmetry breaking are intertwined in this model~\cite{LIU2}. The broken and restored chiral symmetry correspond to a center
symmetric or center asymmetric phases, respectively. Similar studies were developed earlier in~\cite{SHURYAK1,SHURYAK2,TIN}.

In this work we would like to address this interplay between confinement and chiral symmetry breaking
using $N_f$ massless quarks  in the $adjoint$ representation of color   $SU_c(2)$. We will detail the nature of
the flavor symmetry group of the effective action induced by dissociated KvBLL calorons in the confined
phase, and investigate its change into an asymmetric phase at increasing temperature.
Lattice simulations with adjoint quarks~\cite{LATTICEX} have shown that the deconfinement and  restoration of center symmetry occurs well before the restoration of chiral symmetry. These lattice results show that the ratio of the chiral to deconfinement temperatures
is large and decrease with the number of adjoint flavors. More recent lattice simulations have suggested instead
a rapid transition to a conformal phase~\cite{LATTICEZ}.  Effective PNJL models with adjoint fermions have also been
discussed recently~\cite{OGILVIE,KANA}.

The organization of the paper is as follows:
In section 2 we briefly review the index theorem on $S^1\times R^3$  for an adjoint
fermion with twisted boundary condition.  In section 3, we detail the ADHM construction and use it to
derive the anti-periodic adjoint fermion in self-dual BPS dyons.
In section 4, we develop the partition function of an instanton-dyon
ensemble with one light light quark in the adjoint representation of $SU_c(2)$.
By using a series of fermionization and bosonization techniques we construct the
3-dimensional effective action, accommodating  light adjoint quarks
with explicit $SU(N_f)\times Z_{4N_f}$ flavor symmetry.
In section 5, we discuss the nature of the confinement-deconfinement in the quenched
sector ($N_f=0$) of the induced effective action.  In section 6, we show that  for a sufficiently
dense instanton-dyon liquid with light adjoint quarks,
the 3-dimensional ground state is still center symmetric  and  breaks spontaneously
$SU(N_f)\times Z_{4N_f}\rightarrow O(N_f)$ flavor symmetry.  Center symmetry is broken  and chiral symmetry is restored only in  a more dilute instanton-dyon liquid, corresponding to higher temperatures.
Our conclusions are summarized in section 6.  In Appendix A we check that our  ADHM construct reproduces the
expected periodic zero modes for BPS dyons. In Appendix B we derive the pertinent equations for
the anti-periodic adjoint fermions in a BPS monopole without using the ADHM method.
In Appendix C we explicit the ADHM construction for the anti-periodic zero modes in a KvBLL caloron.
In Appendix D  we detail the Fock correction to the mean-field analysis.
In Appendix E we briefly outline the 1-loop analysis for completeness.
In Appendix F we quote
the general result for the 1-loop contribution to the holonomy
potential with $N_f$ adjoint massless quarks.

\section{Index theorem for twisted quarks}

In this section we revisit the general Nye-Singer index theorem for fermions on a finite temperature
Euclidean manifold $S^1\times R^3$ for a general fermion representation. For periodic fermions a very transparent
analysis was  provided by Popitz and Unsal~\cite{POP}. We will extend it to fermions with arbitrary ``twist" (phase),
which is the used for our case of anti-periodic fermions in the adjoint representation.

\subsection{Index}

Consider chiral Dirac fermions on $S^1\times R^3$ interacting with an anti-self-dual gauge field $A$ through

\be
\label{DIR}
\left(D\equiv\gamma_{\mu}D_{\mu}\equiv \gamma_\mu(\partial_{\mu}+igT^{a}A^a_{\mu})\right)\Psi(x)=0
\ee
with twisted fermion boundary conditions ($\beta=1/T$)

\be
\Psi(x_4+\beta,{\bf x})=e^{i\varphi}\Psi(x_4,{\bf x})
\ee
Here $D$ satisfies

\bea
&&D^{\dagger}D=-D_{\mu}D_{\mu}+2\sigma^{m}B_m=DD^{\dagger}+2\sigma^{m}B_m
\eea
For monopoles, the difference between the zero modes of different chiralities in arbitrary R-representation
is captured by  Calias index~\cite{CAL}

\be
{\mathbb I}_{R}=\lim_{M\to 0}\,M\,{\rm Tr} \left<\Psi^\dagger\gamma_5\Psi\right>=
\lim_{M\to 0}{\rm Tr}\left( \gamma_5\frac{M}{-D+M}\right)
\ee
with the Trace carried over spin-color-flavor and space-time. Using the local chiral anomaly condition
for the iso-singlet axial current $J_\mu^5=\Psi^\dagger \gamma_5\gamma_\mu\Psi$ in Euclidean
4-dimensional space

\be
\partial_\mu J_\mu^5=-2M\Psi^\dagger \gamma_5\Psi-\frac {T_R}{8\pi^2}\,F_{\mu\nu}^a{\tilde F}_{\mu\nu}^a
\ee
we can re-write the index in the following form

\be
\label{SURF}
{\mathbb I}_R=
-\frac{1}{2}\int_{S^1\times S^2}d\sigma^2_k\left<J^5_k\right>-\frac{T_R}{16\pi^2}\int_{S^1\times R^3} F^a_{\mu \nu}{\tilde F}^a_{\mu \nu}
\ee
with $T_R$ the Casimir operator in the R-representation.
The second contribution (${\mathbb I}_2$) in (\ref{SURF}) depends only on the
gauge-field, but the first contribution (${\mathbb I}_1$) in (\ref{SURF}) depends on the nature of the fermion field.

\subsection{L, M Dyons}

To evaluate (\ref{SURF}) for  twisted $SU_c(2)$ adjoint fermions in the background of an anti-self-dual or ${\overline M}$ dyon,
we follow Popitz and Unsal~\cite{POP} and write


\bea
\label{SURF1}
&&\left<J^5_k\right>\equiv {\rm Tr}\left<x\left|\gamma^k\gamma _5D\frac{1}{-D^2+M^2}\right|x\right>=\\
&&{\rm Tr}\left<x\left|i\sigma^kD_4\left(\frac{1}{-D^2+M^2+2\sigma \cdot B}-\frac{1}{-D^2+M^2}\right)\right|x\right>\nonumber
\eea
In the ${\overline M}$ anti-dyon background, we have at asymptotic spatial infinity

\bea
\label{SURF2}
-D^2\rightarrow&& -\nabla^2+\left(\left<A_4\right>+\frac{\pi(2p+\frac {\varphi}\pi)}{\beta}\right)^2\nonumber\\
B_m\rightarrow &&-\frac{r_m}{r^3}
\eea
The compact character of $A_4$ on $S^1$ breaks $SU_c(2)\rightarrow {\bf Ab}(SU_c(2))$.
After expanding the ratio with $B$ in (\ref{SURF1}), only the first term carries a non-vanishing net flux in (\ref{SURF})
on $S^2$ thanks to the asymptotic in (\ref{SURF2}). If we recall that the Trace now carries a  summation over the windings along $S^1$
labeled by $p$,  and using the identity

\be
\sum_{p=-\infty}^\infty {\rm sgn}(x+p)=1-2x+2[x]
\ee
we have

\bea
{\mathbb I}_1=&&-\sum_{m=-1}^{m=1}m\sum_{p=-\infty}^{\infty}
{\rm sgn}\left(-\frac{2\pi \nu}{\beta}m+\frac{\pi(2p+\frac \varphi\pi)}{\beta}\right)\nonumber \\
=&&-4\nu+2\left[\nu+\frac{\varphi}{2\pi}\right]-2\left[-\nu+\frac{\varphi}{2\pi}\right]
\eea
For color $SU_c(2)$, $T_R=1/2$ in the fundamental representation and
$T_R=2$ in the adjoint representation. For the latter,

\be
{\mathbb I}_2=-\frac{2}{16\pi^2}\int_{S^1\times R^3} F^a_{\mu \nu}{\tilde F}^a_{\mu \nu}=4\nu
\ee
it follows that

\be
\mathbb I_{M}=\mathbb I_1+\mathbb I_2=2\left[\nu+\frac{\varphi}{2\pi}\right]-2\left[-\nu+\frac{\varphi}{2\pi}\right]
\ee

For the L-dyon we note that  the surface contributions satisfy
$\mathbb I_{1L}=-\mathbb I_{1M}$ since  the asymptotics at spatial infinity have the same $A_4$ with $B_m$
of opposite sign. Therefore we obtain

\be
\mathbb I_{L}=4-\mathbb I_{M}
\ee
whatever the twist $\varphi$ as expected. For anti-periodic fermions with $\varphi=\pi$,
we find that  for $\nu<\frac{1}{2}$, the L-dyon  carries 4 anti-periodic zero modes and the M-dyon
carries 0 zero mode.  For $\frac{1}{2}<\nu<1$, the M-dyon carries  4 zero modes and the L-dyon carries 0 zero mode.
The confining holonomy with  $\nu=\frac{1}{2}$ is special as the zero modes are shared equally
between the $L$- and $M$-dyon, 2 on each. These observations are in agreement with those made
recently in~\cite{KANA}.

\section{ADHM construction of Adjoint zero modes}

In this section we first remind the general framework for the ADHM~\cite{ADHM, MATTIS, KVLL,DPX} construction  for adjoint fermions, and then
 apply it to to the special case of adjoint fermions in the background field of BPS dyons.  
A concise presentation of this approach can be found in~\cite{MATTIS,DPX} whose notations we will use below.
Throughout  this section  we will set  the circle circumference
$\beta=1/T\rightarrow 1$, unless specified otherwise.
We note that our construction is similar in spirit to the one presented in~\cite{ADHMX} for adjoint fermions in calorons,
but is different in some details. In particular,  it does not rely on the replica trick and therefore does not double the size of the ADHM data.

\subsection{ADHM construction}

The basic building block in the ADHM construction is the asymmetric matrix of data $\Delta(x)$ of
dimension $[N+2k]\times [2k]$ for an $SU(N)$ gauge configuration of topological charge $k$. The null
vectors of $\Delta(x)$ can be assembled into a matrix-valued complex matrix $U(x)$ of dimension
$[N+2k]\times [N]$, satisfying $\bar \Delta U=0$ or specifically

\be
\label{H1}
\bar \Delta^{\dot \alpha\lambda}_{i}U_{\lambda u}=0
\ee
with the ADHM label $\lambda=u+i\alpha$ running over $1\leq u\leq N$, $0\leq i\leq k$ and $\alpha,\dot\alpha=1,2$ referring
to the Weyl-Dirac indices which are raised by $\epsilon_2$.  They are orthonormalized by $\bar U U={\bf 1}_N$.  In terms
of (\ref{H1}) the classical ADHM gauge field $A_m$ with $1\leq m\leq 4$  reads

\be
A_m=\bar U\,i\partial_m\,U
\ee
For $k=0$ it is a pure gauge transformation with a field strength $A_{mn}$ that satisfies the self-duality
condition $A_{mn}={}^*A_{mn}$.  For $k\neq 0$ it still satisfies the self-duality condition provided that
\cite{MATTIS}

\be
\label{H3}
\bar \Delta^{\dot \beta\lambda}_{i}\Delta_{\lambda j \dot\alpha}=\delta^{\dot \beta}_{\dot \alpha}f^{-1}_{ij}
\ee
with $f^\dagger=f$ a positive matrix of dimension $[k]\times [k]$. The matrix of data is taken to be linear in the space-time variable $x_n$

\bea
\Delta_{\lambda i \dot\alpha}=&&a_{\lambda i \dot\alpha}+b^{\alpha}_{\lambda i}x_{\alpha\dot \alpha}\nonumber\\
\bar \Delta^{\dot \alpha\lambda}_{i}=&&\bar a^{\dot \alpha\lambda}_{i}+\bar x^{\dot \alpha \alpha}\bar b_{\alpha i}^{\lambda}
\eea
with the quaternionic notation $x_{\alpha\dot \alpha}=x_n(\sigma_n)_{\alpha\dot \alpha}$ and $\sigma_n=({\bf 1}_2, i\vec\sigma)$.

\subsection{Anti-periodic adjoint fermion in general}

Given the matrix of ADHM data as detailed above, the adjoint fermion zero mode  in a self-dual gauge configuration of topological charge $k$ reads~\cite{MATTIS}

\be
\label{HD1}
\lambda_{\alpha}=\bar U Mf\bar b_{\alpha}U-\bar Ub_{\alpha}f\bar M U
\ee
which can be checked to satisfy the Weyl-Dirac equation provided that the Gassmanian matrix $M\equiv M_{\lambda i}$
of dimension $[N+2k]\times [k]$ satisfies the constraint condition

\be
\label{HD2}
\bar \Delta^{\dot \alpha}M+\bar M\Delta^{\dot \alpha}=0
\ee

To unravel the constraints (\ref{H3}) and (\ref{HD2}) it is convenient to re-write the ADHM matrix of data $\Delta(x)$
in quaternionic blocks through a pertinent choice of the complex matrices $a, b$, i.e.

\be
\label{HD3}
\Delta(x)=\left(\begin{array}{cc}
\xi\\
B-x{\bf 1}_2
\end{array}\right)
\ee
with

\bea
&&\xi\equiv \xi_{ui\dot \alpha}\equiv (\xi_{\dot \alpha})_{ui}\nonumber\\
&&B\equiv (B_{\alpha \dot \alpha})_{ij}
\eea
In quaternionic blocks, the null vectors  (\ref{H1}) are

\bea
U(x)\equiv \frac 1{\sqrt{\phi (x)}} \left(\begin{array}{cc}
-{\bf 1}_2\\
u (x)
\end{array}\right)=\frac 1{\sqrt{\phi(x)}} \left(\begin{array}{cc}
-{\bf 1}_2\\
(B^{\dagger}-x^{\dagger}{\bf 1}_2)^{-1}\xi^{\dagger}
\end{array}\right)\nonumber\\
\eea
with the normalization $\phi(x)=1+u^\dagger (x)u(x)$.
To solve the constraint condition (\ref{HD2}) we also define

\be
M\equiv \left(\begin{array}{cc}
c_{uj}\\
M_{\alpha ij}
\end{array}\right)
\ee
and its conjugate $\bar M\equiv \begin{array}{cc}(\bar c_{ju},\bar M^{\alpha}_{ji})\end{array}$. Therefore
the solution to (\ref{HD2}) satisfies $M^{\alpha}=\bar M^{\alpha}$ and the new constraint between the Grassmanians

\be
\label{HD4}
\left [M^{\alpha},B_{\alpha \dot \alpha}\right ]+\bar c \xi_{\dot \alpha }+\bar\xi_{\dot \alpha}c=0
\ee

finally, for periodic gauge configurations on $S^1\times R^3$ such as the  KvBLL calorons or BPS dyons, the
index $k$ is extended to all charges in $\mathbb Z$. It is then more convenient to use the Fourier
representations

\bea
f(z)=&&\sum^\infty_{k=-\infty}f_ke^{i2\pi kz}\nonumber\\
B(z,z^\prime)=&&\sum^\infty_{k,l=-\infty}B_{kl}e^{2\pi i(kz-lz^\prime)}
\eea
which are z-periodic of period 1.

\subsection{Anti-periodic adjoint fermion in a BPS Dyon}

For  BPS dyons the previous arguments apply~\cite{NAHM}. In particular,  for the $SU(2)$ M-dyon
on $S^1\times R^3$, the preceding construct simplifies. In particular, the quaternion blocks in the
ADHM matrix of data in (\ref{HD3}) are simply


\bea
&&\xi=0\nonumber\\
&&B(z,z^\prime)=\delta(z-z^\prime)\frac{1}{2\pi i\nu }\frac{\partial}{\partial z}
\eea
The normalized null vector is readily found in the form

\bea
U=\left(\begin{array}{cc}
0\\
u(x,z)
\end{array}\right)
\eea
with

\bea
u(x,z)=\left({\frac{2\pi vr}{{\rm sinh}(2\pi vr)}}\right)^{\frac 12}e^{i2\pi z v(x_4-i\sigma \cdot x)}
\eea
with the vev $v=\nu/\beta$.

The  constraint (\ref{H3}) following from the self-duality condition  translates to the equation for the
resolvent

\bea
\left(i\frac{\partial}{\partial z}+2\pi \nu x_4\right)^2f(z,z^\prime)+(2\pi \nu r)^2f(z,z^\prime) =\delta(z-z^\prime)\nonumber\\
\eea
The  solution is

\bea
\label{FZZ}
f(z,z^\prime)=&&-\frac{e^{2\pi ivx_4(z-z^\prime)}}{8\pi v r}
\left({\rm sinh}(2\pi vr|z-z^\prime|)\right.\nonumber \\
&&\left. +{\rm coth}(\pi v r) \,{\rm sinh} (2\pi v rz)\, {\rm sinh} (2\pi vr z^\prime)\right.\nonumber\\
&&\left. -{\rm tanh} (\pi v r )\,{\rm cosh}( 2\pi vr z) \, {\rm cosh}( 2\pi vz^\prime)\right)\nonumber\\
\eea
We have explicitly checked that (\ref{FZZ}) satisfies the identities used in the ADHM construction
as noted in~\cite{MATTIS}. In our case these identities read

\bea
&&2\int _{-\frac 12}^{\frac 12}dz_1\tilde f(z,z_1)\left(\frac{\partial }{\partial z_1}\right)\tilde f (z_1,z^\prime)=-(z-z^\prime)\tilde f(z,z^\prime)\nonumber\\
&&-\frac{\partial}{\partial x_i}\tilde f(z,z^\prime)=
2x_i(2\pi\nu )^2\int_{-\frac 12}^{\frac 12 }dz_1\tilde f(z,z_1)\tilde f (z_1,z^\prime)\nonumber\\
\ee
with the definition $f/\tilde f=e^{2\pi ix_4(z-z^\prime)}$, and

\bea
&&\delta(z-z^\prime)-\frac{\partial^2}{\partial z \partial z^\prime}f(z, z^\prime)\nonumber\\
&&-2\pi \nu r\sigma \cdot \hat r\left(\frac {\partial}{\partial z}+\frac{\partial }{\partial {z^\prime}}\right)f(z, z^\prime)-(2\pi \nu r)^2f(z, z^\prime) \nonumber \\
=&&\frac{2\nu \pi r}{\sinh(2\pi \nu r)}(\cosh(2\pi \nu r(z+z^\prime))\nonumber\\
&&+\sigma \cdot \hat r\sinh(2\pi \nu r(z+z^\prime)))
\eea
We  note that the periodicity on $S^1$ translates to the quasi-periodicities

\bea
&&u(x_4+\beta,\vec x, z)=e^{2\pi i\nu z}u(x_4,\vec x, z)\nonumber\\
&&f(x_4+\beta,\vec x,z,z^\prime)=e^{2\pi i\nu (z-z^\prime)}f(x_4,\vec x,z,z^\prime)
\eea

For the adjoint fermion zero-mode, the Grassmanian matrix also simplifies

\be
\label{BP1}
M(z-z^\prime)=M(z^\prime)\,\delta\left(z-z^\prime\pm\frac{1}{2\nu }\right)
\ee
Inserting (\ref{BP1}) in the contraint equation (\ref{HD4})  and noting that now $\xi=0$, yield

\be
\frac{d}{dz}M(z)=0\rightarrow M(z)=M^\pm
\ee
with normalized  constant spinors $M^\pm$. This allows to re-write (\ref{BP1}) in the explicit form

\be
M(z-z^\prime)=M^{+}\delta\left(z-z^\prime+\frac{1}{2\nu }\right)+M^{-}\delta\left(z-z^\prime-\frac{1}{2\nu}\right)\nonumber\\
\ee

With the above in mind, the adjoint zero-mode solution (\ref{HD1}) in the $SU_c(2)$ BPS M-dyon
simplifies to

\bea
\label{ADYON1}
\lambda_{\alpha}^{\pm}(x)=&&-\int^{+\frac 12}_{-\frac 12}dz dz^\prime\, \nonumber\\
&&\times u^{\dagger}(x,z)\epsilon M^{\pm}
f\left(z\mp\frac{1}{2\nu },z^\prime\right)u_{\alpha}(x,z^\prime)\nonumber \\
&&-\int^{+\frac 12}_{-\frac 12} dz dz^\prime\, \nonumber\\
&&\times u^{\dagger}_{\alpha}(x,z)f(z,z^\prime)M^{\pm,T}u\left(x,z^\prime\mp \frac{1}{2\nu }\right)\nonumber\\
\eea
For $\nu>1/2$ the integrations can be undone. 
For that we translate the vectors in (\ref{ADYON1}) to spinors using the quaternionic form
$\lambda_{m}^{\pm}=\lambda_{\alpha ab}^{\pm}\sigma_{mba}$, and make (\ref{ADYON1}) more explicit.
The result is

\bea
\label{SPINOR}
\lambda^\pm_m(x)=&&(f_1(r)\sigma_m+f_2(r)\sigma\cdot \hat r \sigma _m \sigma \cdot \hat r\nonumber \\
&&\pm f_3(r) \sigma_m \sigma \cdot \hat r \pm f_4(r) \sigma \cdot \hat r \sigma_m)\chi
\eea
with $f_{1,2,3,4}$ defined as

\bea
\label{LONG}
&&-16s^2\sinh(s)\cosh (s/2)\sinh(s/2)f_1=\nonumber \\ 
&&s^2 \left(-\left(x^2-1\right)\right) \cosh (s (x-1))+s^2 x^2 \cosh (s (x+1))\nonumber \\ 
&&+2 s^2 x \cosh (s x)-2 s^2 \cosh (s x)-2 s^2 x \cosh (s (x+1))+\nonumber \\ 
&&s^2 \cosh (s (x+1))-s x \sinh (s (x-2))\nonumber\\
&&+2 s \sinh (s (x-1))+s x \sinh (s x)-2 s \sinh (s x)\nonumber\\
&&+\cosh (s (x-2))-\cosh (s x)\nonumber\\ \nonumber\\
&&16s^2\sinh(s)\cosh (s/2)\sinh(s/2)f_2=\nonumber \\
&& (1-2 s^2 (x-1)) \cosh (s x)+s(-s (x^2-1) \cosh (s-s x)\nonumber \\ 
&&+x \sinh (s (x-2))-x \sinh (s x)+2 \sinh (s x)\nonumber \\ 
&&-2 \sinh (s-s x)+s (x-1)^2 \cosh (s (x+1)))\nonumber \\
&&-\cosh (s (x-2)) \nonumber\\ \nonumber\\
&&-16s\sinh(s)\cosh (s/2)\sinh(s/2)f_3=\nonumber \\ 
&& x \cosh (s (x-2))+x (2 s (x-1) \sinh (s)-1) \cosh (s x)\nonumber \\ 
&&-2 s (x-1) (\cosh (s)-1) \sinh (s x) \nonumber \\ \nonumber\\
&&8s\sinh(s)\cosh (s/2)\sinh(s/2)f_4=\nonumber \\
&& \sinh (s x) (s (-x)+\cosh (s) (s (-x)+x \sinh (s)+s)+s)\nonumber \\ 
&&-x \sinh (s) (s (-x)+s+\sinh (s)) \cosh (s x)\nonumber \\ 
\eea
where we have set $s=2\nu\omega_0 r$ and $x=1/2\nu$.
Asymptotically,  the zero modes (\ref{LONG})  simplify to 

\be
f_1\approx -f_2\approx f_3\approx -f_4\rightarrow (2\nu-1)^2e^{\omega_0 (1-2\nu)r}
\ee
and therefore (\ref{SPINOR})  is asymptotically ($r\rightarrow \infty$)

\be
\label{FOUR}
\lambda_m^{\pm }(x)\approx (1\mp \sigma \cdot r)\sigma_m (1\pm \sigma \cdot r)e^{\omega_0(1-2\nu)r}\chi
\ee
We will use this approximation to carry explicitly the analysis below.
The 4 zero modes (\ref{FOUR}) are localized on the M-dyon for $\nu>1/2$, and by duality on the L-dyon for $\nu<1/2$,
in agreement with the index theorem reviewed above.For
$\nu<1/2$ the integration vanishes with $\lambda^+\equiv 0$.

For $\nu>1/2$ we note that  the 4 adjoint
zero modes are normalizable as they fall asymptotically with $e^{\omega_0(1-2\nu)r}$.
(\ref{ADYON1}) can be explicitly checked to be normalized as

\bea
\label{NORM}
&&\int_{R^3} d^3x \,{\rm Tr}\left( \lambda^\pm_\alpha\lambda^{\prime \mp}_\beta\epsilon_{\alpha\beta}\right)
=\frac{1}{8(2\nu\omega_0^2)}\int _{S^2}\nonumber \\
&&\times d\vec S\cdot\vec\nabla\, {\rm Tr}_{z}(\bar M(P+1)M^{\prime}f+\bar M^{\prime}(P+1)Mf)=\nonumber\\
&&\frac{\pi^2(1-\frac{1}{2\nu})}{2(\nu\omega_0 )^3}\bar M^{\prime}\epsilon\bar M
\eea
We note that at $\nu=1/2$ the normalization is upset. This is precisely where the zero modes re-organize
equally between the L- and M-instanton-dyon, a pair on each.

\subsection{Anti-periodic adjoint fermion in a  BPS Dyon with $\nu=\frac 12$}

The case $\nu=1/2$ for the adjoint zero mode is more subtle. The preceding arguments show that
the exponent in it asymptotic decay disapperin this case:
 $e^{\omega_0(1-1)r}=1$. In this limit, the index theorem states that 2 zero modes are localized
on the M-dyon and 2 zero-modes on the L-dyon. In this section, we show that the reduction
of the result (\ref{ADYON1}) for $\nu=1/2$ simplifies. Specifically,

\bea
\label{EQ1}
&&(\lambda^\pm_{\alpha})_{ab}(r)=\frac 1{{\rm sinh}({\omega_0r})}\\
&&\left(\left({\rm cosh}\left(\frac {\omega_0r}2\right)\pm \sigma \cdot \hat r \,{\rm sinh}\left(\frac {\omega_0r}2\right)\right)_{a\beta}\right.\nonumber\\&&\times
(\epsilon M)_{\beta}\,(f({\omega_0r})\pm g({\omega_0r})\sigma \cdot \hat r )_{\alpha b}
\nonumber\\
&&-\epsilon_{\alpha \beta}(f({\omega_0r})\mp g({\omega_0r})\sigma \cdot \hat r)_{a\beta}\,M_{\gamma}\nonumber\\&&\times
\left.\left({\rm cosh}\left(\frac {\omega_0r}2\right)\mp \sigma \cdot \hat r \,{\rm sinh}\left(\frac {\omega_0r}2\right)\right)_{\gamma b}\right)\nonumber
\eea
with

\bea
\label{EQ2}
f({\omega_0r})=&&\frac{1}{4\,{\rm cosh}\left(\frac {\omega_0r}2\right)}(-{\omega_0r}-{\rm sinh} ({\omega_0r})\nonumber)\\
g({\omega_0r})=&&\frac{1}{4\,{\rm sinh}\left( \frac {\omega_0r}2\right)}(-{\omega_0r}+{\rm sinh} ({\omega_0r}))
\eea
(\ref{EQ1}) can be written in a more concise form by translating the vectors to spinors
using the quaternionic form

\be
\lambda_{m}^{\pm}=\lambda_{\alpha ab}^{\pm}\sigma_{mba}
\ee
with

\bea
\label{EQ3}
&&(\lambda^\pm_{\alpha})(r)=\frac 1{{\rm sinh}({\omega_0r})}\nonumber\\
&&\times(f({\omega_0r})\pm g({\omega_0r})\sigma \cdot \hat r )\sigma_m\nonumber\\
&&\times\left({\rm cosh}\left(\frac {\omega_0r}2\right)\pm \sigma \cdot \hat r \,{\rm sinh}\left(\frac {\omega_0r}2\right)\right)
\epsilon M\nonumber\\
&&-\epsilon\left(M^T\left({\rm cosh}\left(\frac {\omega_0r}2\right)\mp \sigma \cdot \hat r \,{\rm sinh}\left(\frac {\omega_0r}2\right)\right)\right.\nonumber\\
&&\left.\times\sigma_m\left(f({\omega_0r})\pm g({\omega_0r})\sigma \cdot \hat r \right)\right)^T
\eea
Using $\sigma_m^T=\epsilon\sigma_m\epsilon$, the transposed of the second term in (\ref{EQ3})
can be reduced. The result is

\bea
\label{EQ4}
&&(\lambda^\pm_{\alpha})(r)=\frac 2{{\rm sinh}({\omega_0r})}\nonumber\\
&&\times(f({\omega_0r})\pm g({\omega_0r})\sigma \cdot \hat r )\sigma_m\nonumber\\
&&\times\left({\rm cosh}\left(\frac {\omega_0r}2\right)\pm \sigma \cdot \hat r \,{\rm sinh}\left(\frac {\omega_0r}2\right)\right)
\chi
\eea
with the identified spinor $\chi=\epsilon M$.
The (color) invariant group norm of (\ref{EQ4}) is  finite. Specifically, if we set
$\lambda_{m, \alpha}^{\pm}=B_{\alpha \beta}^{m\pm}\chi_{\beta}$, then

\bea
\label{EQ5}
&&{\rm Tr}\left(\lambda_{\alpha}^{\pm} \lambda_{\beta}^{\pm}\epsilon_{\alpha \beta}\right)
=\chi^{T}\sum_{m}B^{mT}\epsilon B^{m}\chi \nonumber \\&&=-3\chi^{T}\epsilon \chi
\frac{(f^2(\omega_0 r)-g^2(\omega_0 r))}{{\rm sh}^2(\omega_0 r)}
\eea
which is convergent  in $R^3$.  Note the difference between the Matsubara arrangements
in (\ref{EQ5}) and (\ref{NORM}).
For completeness, we note that (\ref{EQ5}) is the
analogue of the gluino condensate using the anti-periodic zero modes. The periodic zero modes are
briefly discussed in Appendix A  using the same ADHM construct. In Appendix B, we verify explicitly
that the ADHM zero modes are consistent with a direct reduction of the Dirac equation. For completeness,
we detail in Appendix C the ADHM construct for the zero modes around KvBLL instantons.

\section{ Partition function with adjoint fermions}

In this section we will use the adjoint zero modes made explicit
in (\ref{ADYON1}-\ref{FOUR}), to construct the partition function for an ensemble of interacting
dyons and anti-dyons with adjoint fermions. We will show that the partition function is amenable
to a 3-dimensional effective theory. The derivation will be for the non-symmetric case with $\nu>1/2$,
where all the 4 adjoint zero modes are localized on the $M$-dyon (anti-dyon). The non-symmetric case
with $\nu<1/2$ with the adjoint zero modes localized on the $L$-dyon (anti-dyon) is equivalent
and follow by duality $L\leftrightarrow M$ and $\nu\rightarrow\bar\nu= 1-\nu$.   The symmetric case with each $L$ and
$M$ dyons carrying 2 of the 4 adjoint zero modes,  will be understood in the limit $\nu\rightarrow 1/2$.

\subsection{Partition function}

In the semi-classical approximation, the Yang-Mills partition function is assumed to be dominated by an interacting ensemble of
instanton-dyons (anti-dyons). They are constituents of KvBLL instantons (anti-instantons) with fixed holonomy~\cite{KVLL}.
The $SU_c(2)$ grand-partition function   with $N_f$ adjoint Majorana quarks is

\bea
{\cal Z}_{1}[T]&&\equiv \sum_{[K]}\prod_{i_L=1}^{K_L} \prod_{i_M=1}^{K_M} \prod_{i_{\bar L}=1}^{K_{\bar L}} \prod_{i_{\bar M}=1}^{K_{\bar M}}\nonumber\\
&&\times \int\,\frac{f_Ld^3x_{Li_L}}{K_L!}\frac{f_Md^3x_{Mi_M}}{K_M!}
\frac{f_Ld^3y_{{\bar L}i_{\bar L}}}{K_{\bar L}!}\frac{f_Md^3y_{{\bar M}i_{\bar M}}}{K_{\bar M}!}\nonumber\\
&&\times {\rm det}(G[x])\,{\rm det}(G[y])\,\left|{\rm det}\,\tilde{\bf T}(x,y)\right|^{\frac {N_f}2}\nonumber\\
&&\times e^{-V_{D\overline D}(x-y)}e^{-V_L(x-y)}e^{-V_M(x-y)}\nonumber\\
\label{SU2}
\eea
Here $x_{mi}$ and $y_{nj}$ are the 3-dimensional coordinate of the i-dyon of  m-kind
and j-anti-dyon of n-kind. Here
$G[x]$ a $(K_L+K_M)^2$ matrix and $G[y]$ a $(K_{\bar L}+K_{\bar M})^2$ matrix whose explicit form are given in~\cite{DP,DPX}.
The fugacities $f_{i}$ are related to the overall dyon plus anti-dyon density $n_D$~\cite{CALO-LATTICE}.

\begin{figure}[h!]
 \begin{center}
 \includegraphics[width=8cm]{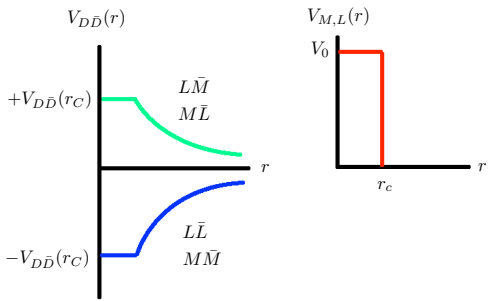}
  \caption{Schematic description for the streamline (left)   and core  (right) potentials between a pair of  $SU_c(2)$
 instanton-dyon and anti-instanton dyon.}
   \label{fig_potdd}
  \end{center}
\end{figure}

$V_{D\bar D}$ is the streamline interaction between ${D=L,M}$ dyons and ${\bar D=\bar L, \bar M}$ anti-dyons as numerically discussed in~\cite{LARSEN,LIU1}. For the $SU(2)$ case  it is Coulombic asymptotically~\cite{LIU1}

\bea
&&V_{D\overline{D}}(x-y)\rightarrow-\frac {C_D}{\alpha_s\, T}\nonumber\\
&&\times\left(\frac 1{|x_M-y_{\overline{M}}|}+\frac 1{|x_L-y_{\overline{L}}|}-\frac 1{|x_M-y_{\overline{L}}|}-\frac 1{|x_L-y_{\overline{M}}|}
\right)\nonumber\\
\label{DDXX}
\eea
The strength of the Coulomb interaction in (\ref{DDXX}) is $C_D=2$.
Following ~\cite{SHURYAK1}, we define the core interactions $V_{L, M}(x-y)$
 between $L\bar L$ and $M\bar M$ respectively, which we assume to
be step functions  of height $V_0$ and range $x_0$

\bea
\label{LMCORE}
V_{M}(x-y)=&&TV_0\,\theta(x_0-2\omega_0\nu|x-y|)\nonumber\\
V_{L}(x-y)=&&TV_0\,\theta(x_0-2\omega_0\bar\nu|x-y|)
\eea
with $x_0/2$ normalized to the dimensionless unit volume

\be
\left(\frac {x_0}2\right)^3=\frac {4\pi}3
\ee
We recall that the $L\bar M$ and $M\bar L$ channels are repulsive.
A  sketch  of the interaction potentials is given  in Fig.~\ref{fig_potdd}.
Below  the core value of $a_{D\bar D}$, the streamline configuration  annihilates
into perturbative gluons.

\subsection{The determinant of the adjoint fermions}

The fermionic determinant in (\ref{SU2}) is composed of all the hoppings between the dyons and anti-dyons
through the adjoint fermionic zero modes.  To explicit the hopping, we consider in details the case $N_f=1$,
and only quote at the end the generalization to arbitrary $N_f$. To explicit the hopping for $N_f=1$, we define

\bea
\label{DD1}
\Psi(x)\equiv &&\sum_{I, \pm}\Psi^{\pm}(x-x_I)\chi^{\pm}_{I}\nonumber\\
\bar \Psi(x)\equiv &&\sum_{\bar J, \pm } \bar \Psi^{\pm}(x-x_{\bar J})\bar \chi^{\pm}_J
\eea
with the sum running over all dyons and anti-dyons, and the 2 Matsubara frequencies
$\pm\, \omega_0$ subsumed in the zero modes. The adjoint dyon and anti-dyon zero modes are labelled by

\be
\label{DD2}
\lambda^\pm_D(x) \equiv \Psi^\pm (x-x_D)\,\chi^\pm_D
\ee
Here $\chi^\pm_D$ is a 2-component Grassmanian spinor and $\Psi^\pm$  a $2\times 2$ valued matrix,
both of which refer to a  D-dyon (anti-dyon). From (\ref{ADYON1}-\ref{FOUR})  the Fourier transforms of $\Psi^\pm$ read

\be
{\tilde\nu}^{-\frac 32}\Psi^+_m(p)=f_1(p)\sigma_m+if_2(p)[\sigma_m,\sigma\cdot \hat p ]+f_3(p)\hat p_m\sigma \cdot \hat p\nonumber\\
\ee
with

\bea
\label{FFF}
&&f_1(p)=\frac{ \tilde \nu }{(p^2+\tilde \nu^2)^2}+\frac{1}{p^3}
\left(\tilde\nu p\frac{(2p^2+\tilde \nu^2)}{(p^2+\tilde \nu^2)^2}-{\rm tan}^{-1}\left(\frac{p}{\tilde\nu}\right)\right)\nonumber\\
&&f_2(p)=\frac{p}{(p^2+\tilde \nu^2)^2}\nonumber\\
&&f_3(p)=-\frac{1}{p^3}\left(\frac{p\tilde \nu(5\tilde p^2+3\tilde \nu^2)}{(p^2+\tilde \nu^2)^2}-3\,{\rm tan}^{-1}\left(\frac{p}{\tilde\nu}\right)\right)
\eea
Here  $p=|\vec p|$ and $\tilde\nu=(2\nu-1)\omega_0$.

In terms of (\ref{DD1}-\ref{DD2}) the hopping action for massive
adjoint quarks  takes the explicit form

\bea
\label{DD3}
&&i\int d^4x (\Psi^T,\bar\Psi ^T)\left(\begin{array}{cc}
m&\epsilon\sigma\cdot \partial\\
-\epsilon \bar \sigma \cdot \partial&m
\end{array}\right)
\left(\begin{array}{cc}
\Psi\\
\bar \Psi
\end{array}\right)\nonumber \\
&&=\sum_\pm (\chi_I^{T\pm},\bar \chi_J^{T\mp})\left(\begin{array}{cc}
im\,\tilde{\bf K}(x_{II^\prime})& {\bf T}^{\pm}(x_{IJ}) \\
-{\bf T}^{T\pm}(x_{IJ})&-im\,\tilde{\bf K}(x_{JJ^\prime})
\end{array}\right)\left(\begin{array}{cc}
\chi^{\pm}_I\\
\bar \chi^{\mp}_J
\end{array}\right)\nonumber\\
\eea
with $x_{IJ}\equiv x_I-x_J$. We note that the matrix entries in (\ref{DD3}) are
$2\times 2$ valued or quaternionic, and that the matrix overall is anti-symmetric
under transposition. This observation  is consistent with the observations made
in~\cite{ADJOINT}.  The matrix entries in (\ref{DD3}) satisfy

\bea
\label{T12XX}
{\bf T}^{\pm}(x_{IJ})=&&-\epsilon \tilde {\bf T}^{\pm }(x_{IJ})\nonumber\\
{\bf T}^{T\pm }(x_{IJ})=&&-\epsilon \tilde {\bf T}^{\mp \dagger}(x_{IJ})\\
\tilde {\bf K}(x_{II^\prime})=&&-\epsilon {\bf K}(x_{II^\prime})
\eea
Using (\ref{T12XX}) we  can rewrite (\ref{SU2}) for massive fermions
in the basis $(\chi^+,\chi^-, \bar\chi^+, \bar\chi^-)^T$ as follows

\begin{eqnarray}
\label{T12}
&&\left|\det \tilde {\bf T}(x,y)\right|^{\frac 12} \nonumber\\
&&\equiv
\left|\det \left(\begin{array}{cccc}
0&-\epsilon m{\bf K}_{ii^\prime}&0&i\epsilon{\bf \tilde T}^+_{ij}\\
-m\epsilon {\bf K}_{ii^\prime}&0&-i\epsilon{\bf \tilde T }^{+\dagger}_{ji}&0\\
0&-i{\bf \tilde T}^{+\star}_{ij}\epsilon&0&m\epsilon {\bf K}_{ii^\prime}\\
i{\bf  \tilde T}^{+ T}_{ji}\epsilon&0&m\epsilon {\bf K}_{ii^\prime}&0
\end{array}\right)\right|^{\frac 12}\nonumber \\
\end{eqnarray}
with dimensionality $4(K_I+K_{\bar I})^2$.
Each of the quaternionic entry in $\tilde{\bf T}_{ij}^+$ is a  ``hopping amplitude" for a fermion between
an instanton-dyon and an instanton-anti-dyon. Each of the quaternion entry in ${\bf K}_{ii^\prime}$ is an overlap
 between two instanton-dyons
or two  instanton-anti-anti-dyons.

\subsection{Hopping amplitudes}

In momentum space
the quaternionic entries are given by

\be
{\bf T}^{\pm}(p)=\Psi^{T\pm}(-p)\epsilon \sigma \cdot p_{\pm} \bar \Psi^{\mp}(p)
\ee
with again $p_\pm =(\pm \omega_0, \vec p)$.
Since

\be
\Psi^T(p)=\epsilon \Psi(p)\epsilon
\ee
we also have the identities

\bea
\label{T12X}
{\bf T^{\pm}}(p)=&&-\epsilon\Psi^{\pm}(-p)(\pm \omega_0+i\sigma\cdot p)\bar \Psi^{\mp}(p)\nonumber\\
{\bf T}^{T\pm}(p)=&&-\epsilon \bar\Psi^{\mp}(p) (\mp \omega_0+i\sigma \cdot p) \Psi^{\pm}(-p)
\eea
We note the relations

\bea
\Psi^{\pm}(p)=&&\bar \Psi^{\mp}(p)\nonumber\\
(\Psi^{\pm})^{\dagger}(-p)=&&\Psi^{\mp}(p)
\eea
and therefore we have the additional identities

\bea
{\bf T}^{\pm}(p)=&&-\epsilon \tilde  {\bf T}^{\pm}(p)\nonumber\\
{\bf T}^{T\pm}(p)=&&-\epsilon\tilde  {\bf T}^{\mp ^\dagger}(-p)
\eea
Here, we have

\be
\tilde {\bf T}^+(p)=\Psi^{+}(-p)(\omega_0+i\sigma \cdot p)\Psi^{+}(p)
\ee
or more explicitly

\bea
\label{T++}
&&\omega_0^{3}{\tilde\nu}^{-3}\tilde {\bf T}^+(p)=\nonumber\\
&&\left(3f_1^2+f_3^2+2f_1f_3-8f_2^2+8f_1f_2\frac{p}{\omega_0}\right)\omega_0\nonumber \\
&&+i\sigma \cdot p\left(-f_1^2+f_3^2+2f_1f_3+8f_2^2+8f_1f_2\frac{\omega_0}{p}\right)
\eea
We also have

\bea
{\bf K}(p)=&&\Psi^{-}(-p)\Psi^{+}(p)=\Psi^{+\dagger}(p)\Psi^+(p)\nonumber \\
=&&\omega_0^{-3}\tilde \nu^{3}\left(3f_1^2+f_3^2+2f_1f_3+8f_2^2\right)
\eea

\section{Effective action without adjoint fermions}

In this section we will derive the 3-dimensional effective action in the case
without the adjoint fermions,to be referred to as $N_f=0$ case below.
We will analyze it in the limit of weak coupling and large densities across the transition region. We will
explicitly derive the induced effective potential for the $SU_c(2)$ holonomies $\nu, \bar\nu$ and show
that for a critical density the ground state of the 3-dimensional effective theory confines.

\subsection{Bosonic fields}

Following~\cite{DP,LIU1,LIU2} the moduli
determinants in (\ref{SU2}) can be fermionized using 4 pairs of ghost fields $\chi^\dagger_{L,M},\chi_{L,M}$ for the dyons
and 4 pairs of ghost fields $\chi^\dagger_{{\bar L},{\bar M}},\chi_{{\bar L},{\bar M}}$ for the anti-dyons. The ensuing Coulomb factors from the determinants are then bosonized using 4 boson fields $v_{L,M},w_{L,M}$ for the dyons and similarly for
the anti-dyons.  The result is

\bea
&&S_{1F}[\chi,v,w]=-\frac {T}{4\pi}\int d^3x\nonumber\\
&&\left(|\nabla\chi_L|^2+|\nabla\chi_M|^2+\nabla v_L\cdot \nabla w_L+\nabla v_M\cdot \nabla w_M\right)+\nonumber\\
&&\left(|\nabla\chi_{\bar L}|^2+|\nabla\chi_{\bar M}|^2+\nabla v_{\bar L}\cdot \nabla w_{\bar L}+\nabla v_{\bar M}\cdot \nabla w_{\bar M}\right)
\label{FREE1}
\eea

For the interaction part $V_{D\bar D}$, we note that
the pair Coulomb interaction in (\ref{SU2}) between the dyons and anti-dyons can also be bosonized using
standard methods~\cite{POLYAKOV,KACIR,ALL}  in terms of $\sigma$ and $b$ fields.   As a result each dyon species acquire additional
fugacity factors such that

\be
M:e^{-b-i\sigma}\,\,\,\,\, L:e^{b+i\sigma}\,\,\,\,\, \bar M: e^{-b+i\sigma}\,\,\,\,\, \bar L :e^{b-i\sigma}
 \ee
with an additional contribution to the free part (\ref{FREE1})

\bea
&&S_{2F}[\sigma, b]=T\int d^3x\, d^3y\\
&&\times\left(b(x)V^{-1}(x-y) b(y)+ \sigma(x)V^{-1}(x-y)\sigma(y)\right)\nonumber
\label{FREE2}
\eea
The streamline interaction is asymptotically Coulombic and attractive in the $L\bar L$ and $M\bar M$ channels
with

\be
V(r)\approx -\frac{C_D}{\alpha_s}\frac 1{Tr}=-\frac{2}{\alpha_s}\frac 1{Tr}
\ee
and repulsive in the $\bar L M$ and $L\bar M$ channels as illustrated in Fig.~\ref{fig_potdd}. At short distances,
these 4-channels reduce to perturbative gluons that should be subtracted. We follow~\cite{TIN} and introduce
a core interaction  as illustrated in Fig.~\ref{fig_potdd}  to achieve that. Specifically, for the core interactions $V_{L,M}(r)$,
we have

\bea
S_{3F}[\phi_1, \phi_2]=\int d^3x \left(\phi_1^{\dagger}V_M^{-1}\phi_1+\phi_2^{\dagger}V_{L}^{-1}\phi_2\right)
\eea
and the interaction part is now

\bea
&&S_I[v,w,b,\sigma,\chi]=-\int d^3x \nonumber\\
&&f_M\left(4\pi v_m+|\chi_M    -\chi_L|^2+v_M-v_L\right)\nonumber\\
&&\times e^{-b+i\sigma+i\phi_1^\dagger}e^{w_M-w_L}+\nonumber\\
&&f_L\left(4\pi v_l+|\chi_L    -\chi_M|^2+v_L-v_M\right)\nonumber\\
&&\times e^{+b-i\sigma+i\phi_2^\dagger}e^{w_L-w_M}+\nonumber\\
&&f_{\bar M}\left(4\pi v_{\bar m}+|\chi_{\bar M}    -\chi_{\bar L}|^2+v_{\bar M}-v_{\bar L}\right)\nonumber\\
&&\times e^{-b-i\sigma+i\phi_1}e^{w_{\bar M}-w_{\bar L}}+\nonumber\\
&&f_{\bar L}\left(4\pi v_{\bar l}+|\chi_{\bar L}    -\chi_{\bar M}|^2+v_{\bar L}-v_{\bar M}\right)\nonumber\\
&&\times e^{-b-i\sigma+i\phi_2}e^{w_{\bar L}-w_{\bar M}}
\label{FREE3}
\eea
without the fermions. The minimal modifications to (\ref{FREE3})  due to the hopping fermions in the adjoint representation
will be detailed below.

In terms of (\ref{FREE1}-\ref{FREE3}) the instanton-dyon partition function (\ref{SU2}) can be exactly re-written as an interacting
effective field theory in 3-dimensions,

\bea
{\cal Z}_{D\overline D}[T]\equiv \int &&D[\chi]\,D[v]\,D[w]\,D[\sigma]\,D[b]\,D[\phi]\nonumber\\
&&\times e^{-S_{1F}-S_{2F}-S_{3F}-S_{I}}
\label{ZDDEFF}
\eea
In the absence of the fields $\sigma, b, \phi$ (\ref{ZDDEFF}) reduces to the 3-dimensional effective field theory
discussed in~\cite{DP} which was found to be integrable. In the presence of $\sigma, b, \phi$ the integrability is lost as the
dyon-anti-dyon screening upsets the hyper-Kahler nature of the moduli space.
Since the effective action in (\ref{ZDDEFF}) is linear in the $v_{M,L,\bar M,\bar L}$, the latters   are auxiliary fields that
integrate  into delta-function constraints. However and for convenience, it is best to shift away
the $b,\sigma$ fields from (\ref{FREE3}) through

\be
&&w_M-b+i\sigma\rightarrow w_M\nonumber\\
&&w_{\bar M}-b-i\sigma\rightarrow w_{\bar M}
\label{SHIFT}
\ee
which carries unit Jacobian and no anomalies, and recover them in the pertinent arguments of the delta function constraints as

\bea
&&-\frac{T}{4\pi}\nabla^2w_M+f  e^{i\phi_1^\dagger}e^{w}-f e^{i\phi_2^\dagger}e^{-w}\nonumber\\
&&=\frac {T}{4\pi}\nabla^2(b-i\sigma)\nonumber\\
&&-\frac{T}{4\pi}\nabla^2w_L+fe^{i\phi_2^\dagger} e^{-w}-f e^{w}=0
\label{DELTA}
\ee
with $w\equiv w_M-w_L$, $f\equiv \sqrt{f_Mf_L}$, and similarly for the anti-dyons.

\subsection{ Effective action with $N_f=0$}

In~\cite{DP} it was observed that the classical solutions
to (\ref{DELTA}) can be used to integrate the $w^\prime$s in (\ref{ZDDEFF}) to one loop. The resulting bosonic determinant
was shown to cancel against the fermionic determinant after also integrating over the $\chi^\prime$s in (\ref{ZDDEFF}). This
holds for our case as well. However, the presence of $\sigma, b, \phi$ makes the additional parts of (\ref{ZDDEFF})  still very
involved in 3 dimensions. To proceed further, we solve the constraint (\ref{DELTA})

\bea
\label{DELTASOLVED}
&&b-i\sigma=w+\frac{8\pi f}{T(-\nabla^2+M_D^2)}(e^{i\phi_1^{\dagger}}e^{w}-e^{i\phi_2^{\dagger}}e^{-w})\nonumber\\
&&b+i\sigma=\bar w+\frac{8\pi f}{T(-\nabla^2+M_D^2)}(e^{i\phi_1}e^{\bar w}-e^{i\phi_2}e^{-\bar w})\nonumber\\
\eea
with a screening mass $M_D$ to be fixed variationally. In terms of (\ref{DELTASOLVED}),
the effective action without the fermionic contributions ($N_f=0$)  is,

\bea
&&S=S_{\phi}+T\bar wV^{-1}w+(-4\pi f\nu (e^{w}e^{i\phi_1^{\dagger}}+e^{\bar w}e^{i\phi_1})\nonumber \\&&+8\pi f(e^{w}e^{i\phi_1^{\dagger}}\frac{V^{-1}}{M_D^2+\nabla^2}\bar w+e^{i\phi_1}e^{\bar w}\frac{V^{-1}}{-\nabla^2+M_D^2}w))\nonumber \\&&+(-4\pi f\bar \nu (e^{-w}e^{i\phi_2^{\dagger}}+e^{-\bar w}e^{i\phi_2})\nonumber \\&&-8\pi f(e^{-w}e^{i\phi_2^{\dagger}}\frac{V^{-1}}{-\nabla^2+M_D^2}\bar w+e^{-\bar w}e^{i\phi_2}\frac{V^{-1}}{M_D^2-\nabla^2}w)\nonumber  \\&&+\frac{(8\pi f)^2}{T}(e^{i\phi_1^{\dagger}}e^w-e^{i\phi_2^{\dagger}}e^{-w})\frac{1}{M_D^2-\nabla^2}V^{-1}\nonumber \\ &&\times \frac{1}{M_D^2-\nabla^2}(e^{i\phi_1}e^{\bar w}-e^{i\phi_2}e^{-\bar w})\nonumber \\&&+{\rm Tr} \ln(1+\frac{8\pi  f}{T(M_D^2-\nabla^2)}(e^{i\phi_1^{\dagger}}e^w+e^{i\phi_2^{\dagger}}e^{-w}))\nonumber \\&&+{\rm Tr} \ln(1+\frac{8\pi  f}{T(M_D^2-\nabla^2)}(e^{i\phi_1}e^{\bar w}+e^{i\phi_2e^{-\bar w}}))
\eea
with $v_l=v_{\bar l}=\nu$ and $v_m=v_{\bar m}=\bar\nu=1-\nu$. Thus, for constant $w$ we have

\bea
&&S=S_{\phi}+V_3C_{D}\alpha_s \bar w wM_D^2\nonumber \\&&+\int 4\pi f(-\nu(e^{i\phi_1^{\dagger}}e^w+e^{i\phi_1}e^{\bar w})+\nonumber \\&&+2C_D\alpha_s(we^{i\phi_1}e^{\bar w}+\bar we^{i\phi_1^{\dagger}}e^{w}))\nonumber \\&&+\int 4\pi f (-\bar \nu(e^{i\phi_2^{\dagger}}e^{-w}+e^{i\phi_2}e^{-\bar w})\nonumber \\&&-2C_D\alpha_s(we^{i\phi_2}e^{-\bar w}+\bar we^{i\phi_2^{\dagger}}e^{-w}))\nonumber \\&&+\frac{(8\pi f)^2}{T}(e^{i\phi_1^{\dagger}}e^w-e^{i\phi_2^{\dagger}}e^{-w})\frac{1}{M_D^2-\nabla^2}V^{-1}\nonumber \\&& \times \frac{1}{M_D^2-\nabla^2}(e^{i\phi_1}e^{\bar w}-e^{i\phi_2}e^{-\bar w})\nonumber \\&&+{\rm Tr} \ln\left(1+\frac{8\pi  f}{T(M_D^2-\nabla^2)}(e^{i\phi_1^{\dagger}}e^w+e^{i\phi_2^{\dagger}}e^{-w})\right)\nonumber \\&&+{\rm Tr} \ln\left(1+\frac{8\pi  f}{T(M_D^2-\nabla^2)}(e^{i\phi_1}e^{\bar w}+e^{i\phi_2}e^{-\bar w})\right)\nonumber\\
\eea
To proceed further, we will treat the core interaction using the cumulant expansion. In leading order,
only the second cumulant is retained, and the result is

\bea
\label{XZX}
&&\frac{\ln Z}{V_3}\approx \nonumber \\ &&+T\alpha_sC_DM_D^2\left(\bar w+\frac{16\pi f}{TM_D^2}\sinh \bar w\right)\nonumber\\&&\times\left(w+\frac{16\pi f}{TM_D^2}\sinh w\right)\nonumber \\&&-4\pi f(\nu(e^w+e^{\bar w})+\bar \nu (e^{-w}+e^{-\bar w}))\nonumber \\&&+\int d^3r (e^{-V_1}-1)F_1+\int d^3 r(e^{-V_2}-1)F_2 \nonumber \\&&+\int \frac{d^3p}{(2\pi)^3}\ln\left(1+\frac{8\pi f}{T}\frac{e^{w}+e^{- w}}{M_D^2+p^2}\right)\nonumber \\&&+\int \frac{d^3p}{(2\pi)^3}\ln\left(1+\frac{8\pi f}{T}\frac{e^{\bar w}+e^{-\bar w}}{M_D^2+p^2}\right)
\eea
with $F_2=F_1(w\rightarrow -w)$ and

\bea
F_1= &&16\pi^2 f^2e^{w+\bar w}\nonumber\\
&&\times \left|-\nu +2C_D\alpha_s\bar w
+\int \frac{2}{T}\frac{d^3p}{(2\pi)^3}\frac{1}{p^2+M_D^2}\right|^2\nonumber \\&&+\frac{(8\pi f)^2}{T}e^{w+\bar w}
\nonumber\\&&\times\int d^3 r_1d^3r_2 G_{M_D}(r-r_1)V^{-1}(r_1-r_2)G_{M_D}(r_2)\nonumber\\
\eea

\subsection{Effective potential with $N_f=0$}

For small $\alpha_s$ and strong screening,  we may neglect the terms proportional to $\alpha_s$
and drop the  screening contributions. Since $\bar w=w^{\dagger}$,
the effective potential associated to (\ref{XZX}) and including the 1-loop perturbative contribution
for finite holonomy is

\bea
\label{XPX}
-\frac{{\cal P}_D}{8\pi f}=&&-\cos{\sigma} \,(\nu e^{ {\bf b}}+\bar \nu e^{-{\bf b}})
\nonumber \\&&+n\left(\frac{e^{2{\bf b}}}\nu +\frac{e^{-2{\bf b}}}{\bar \nu}\right)\nonumber \\
&&+\frac {4\pi^2}3 \frac{T^3}{8\pi f}\nu^2{\bar\nu}^2
\eea
with ${\bf b}={\rm Re}w$ and

\bea
&&n=\frac{2\pi f(1-e^{-V_0})}{(2\pi T/x_0)^3} \equiv
\frac{2\pi f}{T^3}\frac{32}{3\pi^2}(1-e^{-V_0})
\eea
The extremum in $\sigma\equiv {\rm Im}w$ in (\ref{XPX}) occurs at  $\sigma=0$. The minimum  with respect to ${\bf b}$
is fixed by the quartic equation for $e^{\bf b}$

\be
2n \left(\frac{e^{2{\bf b}(\nu)}}\nu -\frac{e^{-2{\bf b}(\nu)}}{\bar \nu}\right)=(\nu e^{ {\bf b}(\nu)}-\bar \nu e^{-{\bf b}(\nu)})
\label{BMIN}
\ee
with  ${\bf b}(\nu)$ as a solution.  (\ref{BMIN}) admits always the symmetric solution ${\bf b}(1/2)=0$ as an explicit solution
for large $n$. The quenched effective  potential for the holonomy with $N_f=0$ follows in the form

\bea
\label{XPXX}
-\frac{{\cal P}_D}{8\pi f}\rightarrow &&-(\nu e^{ {\bf b}(\nu)}+\bar \nu e^{-{\bf b}(\nu)})
\nonumber \\&&+n\left(\frac{e^{2{\bf b}(\nu)}}\nu +\frac{e^{-2{\bf b}(\nu)}}{\bar \nu}\right)\nonumber \\
&&+\frac {4\pi^2}3 \frac{T^3}{8\pi f}\nu^2{\bar\nu}^2
\eea

\begin{figure}[h!]
 \begin{center}
  \includegraphics[width=6cm]{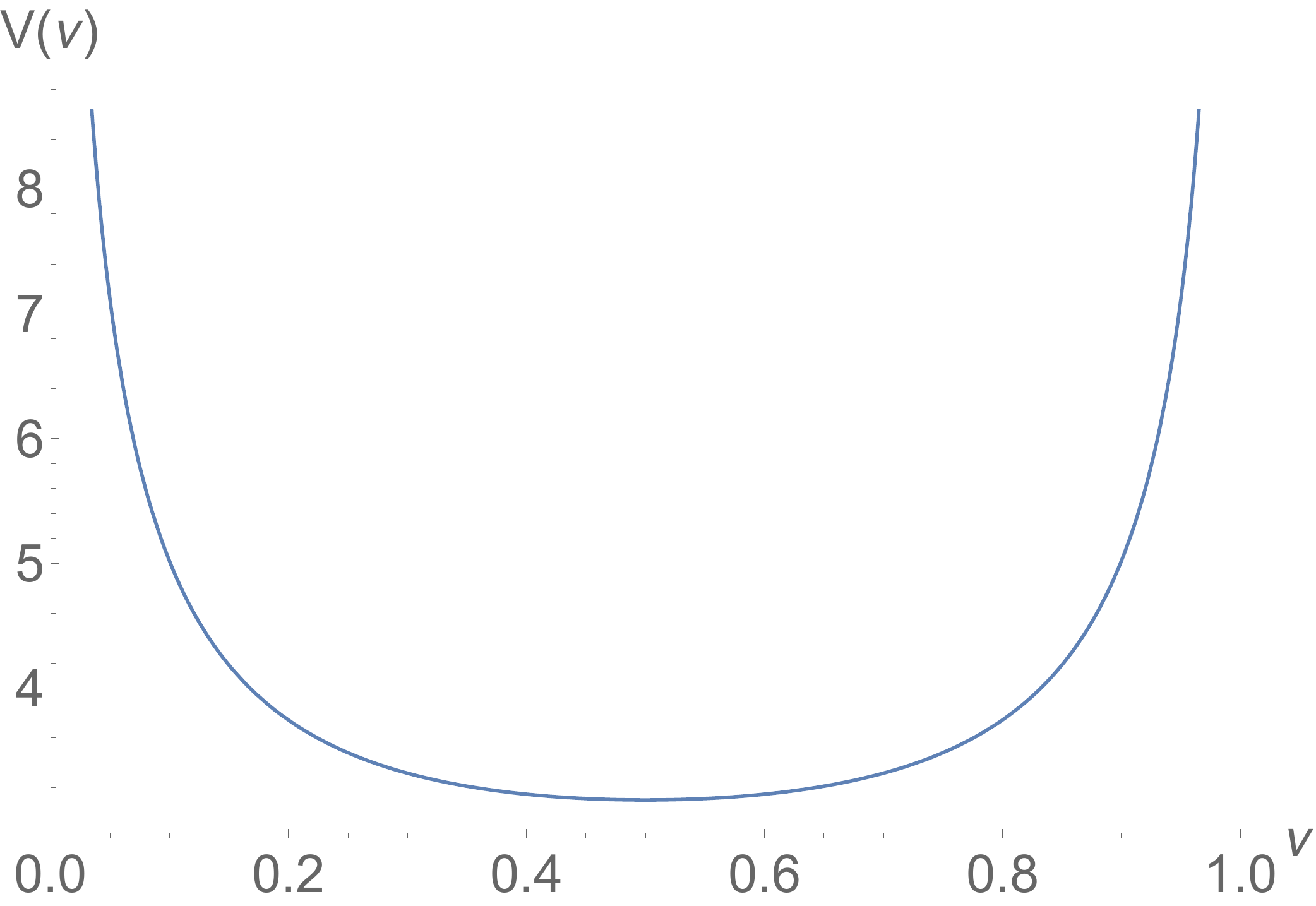}
    \includegraphics[width=6cm]{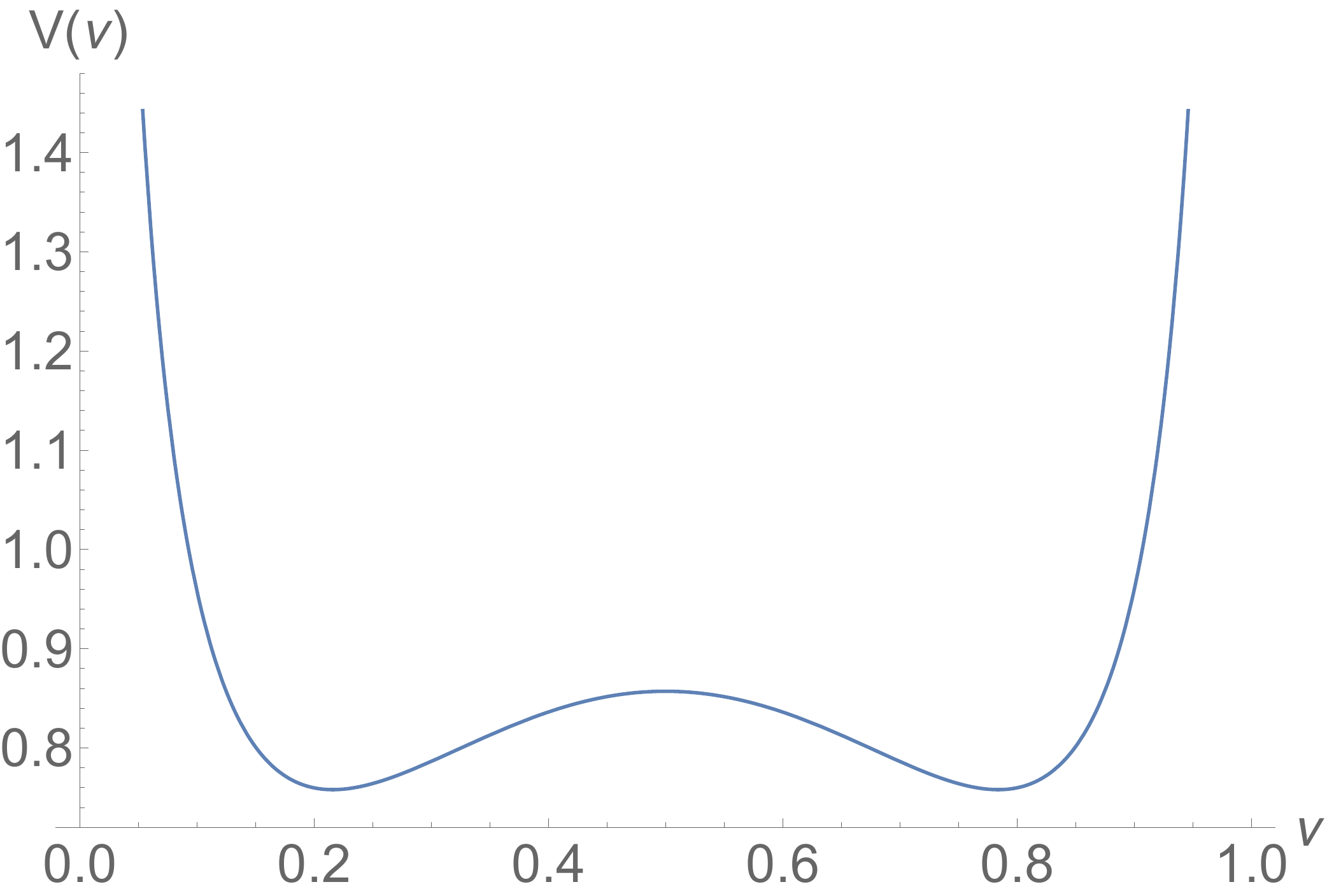}
   \caption{The holonomy potential (\ref{XPXXX}) for  the density $n=1$, in a ``symmetric phase" (above), compared to
its shape at smaller density $n=0.4$, in an    ``asymmetric phase" (below).}
    \label{fig_V1}
  \end{center}
\end{figure}

We note that (\ref{XPXX}) is similar but not identical to the effective potential
discussed in~\cite{SHURYAK1}) using an excluded volume approach.
(\ref{XPXX}) admits a critical instanton-dyon density $n_C$ above which
the minimum of the quenched  potential (\ref{XPXX}) occurs for $\nu=1/2$ or in the confined phase,
and below which two minima develop moving away from $\nu=1/2$ towards the $\nu=0, 1$ or deconfined phase.
To proceed further, we fix  ${V_0}={\rm ln}2$  with $n\approx {\pi f}/{T^3}$.
(\ref{XPXX}) reduces to

\be
\label{XPXXX}
-\frac{{\cal P}_D}{8\pi f}\rightarrow &&n\left(\frac{e^{{2\bf b}(\nu)}}{\nu}+\frac{e^{{-2\bf b}(\nu)}}{\bar \nu}\right)\nonumber\\
&&-\left(\nu e^{{\bf b}(\nu)}+\bar \nu e^{{-\bf b}(\nu)}\right)+\frac{\pi^2}{6n}\nu^2\bar\nu^2
\ee
as shown in Fig.~\ref{fig_V1} for $n=1$  and Fig.~\ref{fig_V05} for
$n=0.4$.  The critical density is found numerically to be $n_D\approx 0.56$  or ${8\pi f}/{T^3}\approx 4.48$.
For $n<n_C$, (\ref{XPXXX}) displays two minima at  $\nu_1<1/2$ and $\nu_2=1-\nu_1$.
For $n>n_C$, we have a single minimum at $\nu=1/2$. The alternative choice of
the core  ${V_0}\rightarrow {\nu V_0}$,  yields a finite effective potential at $\nu=0, 1$.
For $\nu V_0=2\nu$, the critical density occurs at a larger density with $n_C\approx 3.7$,
and  a minimum at ${\bf b}=0$ for $n>n_C$,

\be
-\frac{{\cal P}_{D{\rm min}}}{8\pi f}=4n-1+\frac{\pi^2}{96n}
\ee

%
%
%

\subsection{Electric and magnetic masses with $N_f=0$}

In the center symmetric phase with $\nu=1/2$ with $N_f=0$,  we may define a class of electric and magnetic masses
as the curvatures of the induced potential $-{\cal P}_D$~\cite{SHURYAK1}. Specifically, we have

\bea
Tm_{E}^2=&&\frac{1}{2C_{D}\alpha_s}\frac{\partial^2(-{\cal P}_D)}{\partial^2{\bf  b}}\nonumber \\=&&\frac{4nT^3}{\alpha_sC_D}(8n-1)\nonumber\\
Tm_{M}^2=&&\frac{1}{2C_{D}\alpha_s}\frac{\partial^2(-{\cal P}_D)}{\partial^2{\sigma}}\nonumber\\ =&&\frac{4nT^3}{\alpha_sC_D}
\eea
We note that $M_E^2/M_M^2=8n-1>1$ in  the symmetric phase since $n_D\approx 0.56>1/4$. These masses are distinct
from the electric and magnetic screening masses $M_{E,M}$ following from the decorrelation of the electric and magnetic fields in the
instanton-dyon liquid as discussed in~\cite{LIU1}. The latters are space-like poles in suitably defined propagators.

\section{Effective action with adjoint fermions}


\subsection{Fermionic fields with $N_f=1$}

  To fermionize the determinant(\ref{T12})
and for simplicity, consider first the case of $N_f=1$ flavor an the lowest 2 Matsubara frequencies $\pm\omega_0$.
As we noted earlier, the quaternionic matrix in (\ref{T12}) is real and anti-symmetric of dimensionality $4(K_I+K_{\bar I})^2$.
Its fermionization will only require the use of a single species of
Grassmanians with no need for their conjugate.  Specifically, we have

\be
\left|{\rm det}\,\tilde{\bf T}\right|^{\frac 12} =\int   D[\chi]\,\, e^{\, \chi^{T}  {\bf \tilde T} \, \chi}
\label{TDET}
\ee
 with $ \chi=(\chi^{+},\chi^{-},\bar \chi^{+},\bar \chi^{-})$. This is the analogue
 of the Majorana-like representation for our hopping matrix in Euclidean $S^1\times R^3$.
We can re-arrange the exponent in (\ref{TDET}) by defining  a Grassmanian source
${\mathbb J}(x)=({\mathbb J}^{+}(x),{\mathbb J}^{-}(x), \bar {\mathbb J}^{+}(x),\bar {\mathbb J}^{-}(x))^T$ with

\be
\mathbb J^{+}_\alpha(x)=\sum^{K_L}_{I=1}\chi^{+I}_{\alpha}\delta^3(x-x_{I})\nonumber\\
\bar {\mathbb J}^{+\dot\beta}(x)=\sum^{K_{\bar I}}_{J=1} \bar \chi^{+\dot \beta }_{2J}\delta^3(x-y_{J})
\label{JJ}
\ee
and by introducing 2 additional fermionic fields  $ \psi(x)=(\psi_{+}(x),\psi_{-}(x),\bar \psi_{+},\bar \psi_{-})^T$. Thus

\be
e^{\,\chi^{T}  \tilde{\bf T}\,\chi}=\frac{\int D[\psi]\,{\rm exp}\,(-\int\psi^{T} \tilde {\bf G}\, \psi +
2\int {\mathbb J}^{T} \psi)}{\int d
D[\psi]\, {\rm exp}\,(-\int \psi^T \tilde {\bf G} \,\psi) }
\label{REFERMIONIZE}
\ee
with $\tilde{\bf G}$ a $4\times 4$ chiral block matrix defined by:
\be
\tilde {\bf G}\tilde {\bf T}={\bf 1}
\ee
Fo massless adjoint quarks, we have the explicit form

\begin{eqnarray}
\left(\begin{array}{cccc}
0&0&0&i\epsilon{\bf G}^{T}(y-x)\\
0&0&-i\epsilon {\bf G}^{\star}(x-y)&0\\
0&-i{\bf G}^{\dagger}(y-x)\epsilon&0&0\\
i{\bf G}(x-y)\epsilon&0&0&0
\end{array}\right)
\label{GG}\nonumber \\
\end{eqnarray}
with entries ${\bf T}{\bf  G}={\bf 1}$. The Grassmanian source contributions in (\ref{REFERMIONIZE}) generates a string
of independent exponents for the instanton-dyons and instanton-anti-dyons

\begin{eqnarray}
\prod^{K_I}_{I=1}e^{2\chi^{+T}_{I}\psi_{+}(x_I)+2\chi^{-T}_I\psi_{-}(x_I)}\nonumber \\ \times
\prod^{K_{\bar I}}_{J=1}e^{2\bar \chi^{+T}_J \bar \psi_{+}(y_J)+2\bar \chi ^{-T}_J\bar \psi_{-}(y_J)}
\label{FACTOR}
\end{eqnarray}
The Grassmanian integration over the $\chi_i$ in each factor in (\ref{FACTOR}) is now readily done to yield

\be
\prod_{I}\left[\psi^{T}_{+}\epsilon \psi_{+}\psi^{T}_{-}\epsilon\psi_{-}\right]\prod_J\left[\bar \psi^{T}_{+}\epsilon \bar \psi_{+}\bar \psi^{T}_{-}\epsilon \bar \psi_{-}\right]\nonumber \\=\prod_{I}\left[\psi^{T}_{+}\epsilon \psi_{-}\psi^{T}_{+}\epsilon\psi_{-}\right]\prod_J\left[\bar \psi^{T}_{+}\epsilon \bar \psi_{-}\bar \psi^{T}_{+}\epsilon \bar \psi_{-}\right]
\label{PLPR}
\ee
for the instanton-dyons and instanton-anti-dyons.
The net effect of the additional fermionic determinant in (\ref{SU2}) is to shift the dyon
and anti-dyon fugacities in (\ref{FREE3}) through

\bea
f_I\rightarrow &&f_I\, \psi^{T}_{+}\epsilon  \psi_{-}(x_I)\psi^{T}_{+}\epsilon  \psi_{-}(x_I)\nonumber\\
f_{\bar I}\rightarrow &&f_{\bar I}\, \bar \psi^{T}_{+}\epsilon \bar \psi_{-}(x_{\bar I})\bar \psi^{T}_{+}\epsilon \bar \psi_{-}(x_{\bar I})
\label{SUB}
\eea

\subsection{Resolving the constraints}

In terms of (\ref{FREE1}-\ref{FREE3})  and the substitution
(\ref{SUB}), the dyon-anti-dyon partition function (\ref{SU2})
for finite $N_f$ can be exactly re-written as an interacting
effective field theory in 3-dimensions,

\bea
{\cal Z}_{1}[T]\equiv &&\int D[\psi]\,D[\chi]\,D[v]\,D[w]\,D[\sigma]\,D[b]\,D[\phi_1]\,D[\phi_2]\nonumber\\&&\times
e^{-S_{1F}-S_{2F}-S_{I}-S_\psi-S_{\phi}}
\label{ZDDEFF}
\eea
with the additional $N_f=1$ chiral fermionic contribution $S_\psi=\psi^{T}\tilde{\bf G}\,\psi$.
Since the effective action in (\ref{ZDDEFF}) is linear in the $v_{M,L,\bar M,\bar L}$, the latters
integrate to give the following constraints

\bea
&&-\frac{T}{4\pi}\nabla^2w_M+(\psi^{T}_{+}\epsilon \psi_{-})^2f_M  e^{w_M-w_L+i\phi_1^{\dagger}}\nonumber\\&&
-f_L\e^{w_L-w_M+i\phi_2^\dagger}=\frac {T}{4\pi}\nabla^2(b-i\sigma)\nonumber\\
&&-\frac{T}{4\pi}\nabla^2w_L- (\bar \psi^{T}_{+}\epsilon \bar \psi_{-})^2f_Me^{w_M-w_L+i\phi_1^{\dagger}}
\nonumber\\&&
+f_Le^{w_L-w_M+i\phi_2^{\dagger}}=0
\label{DELTA}
\eea
and similarly for the anti-dyons with $M,L, \psi\rightarrow \overline M, \overline L, \bar\psi$.
To proceed further the formal classical solutions to the constraint equations or $w_{M,L}[\sigma, b]$
should be inserted back into the 3-dimensional effective action. The result is

\bea
{\cal Z}_{1}[T]=\int D[\psi]\,D[\sigma]\,D[b]\,D[\phi]e^{-S}
\label{ZDDEFF1}
\eea
with the 3-dimensional effective action

\bea
S=&&S_F[\sigma, b]+S[\phi]+\int d^3x\,\psi^T \tilde{\bf G} \psi\\
&&-4\pi f_M v_m\int d^3x\,( \psi^{T}_{+}\epsilon \psi_{-})^2 e^{w_M-w_L+i\phi_1^{\dagger}} \nonumber \\
&&-4\pi f_M v_m\int d^3x\,(\bar \psi^{T}_{+}\epsilon \bar \psi_{-})^2e^{w_{\bar M}-w_{\bar L}+i\phi_1} \nonumber \\
&& -4\pi f_Lv_l\int d^3x\,( e^{w_{L}-w_{M}+i\phi_2^{\dagger}}+e^{w_{ \bar L}-w_{ \bar M}+i\phi_2} )\nonumber \\
\label{NEWS}
\eea
Here $S_F$ is $S_{2F}$ in (\ref{FREE2}) plus additional contributions resulting from the $w_{M,L}(\sigma, b)$ solutions
to the constraint equations (\ref{DELTA}) after their insertion back.  The fermionic contributions in (\ref{NEWS}) are
$Z_4$ symmetric.

\subsection{Ground state with $N_f=1$}

We first consider  the massless case with
$m=0$. The uniform ground state of the 3-dimensional effective theory described by (\ref{ZDDEFF}-\ref{NEWS})
corresponds to $b, \sigma, w$ constant, with a finite condensate
with

\bea
\label{HARTREE}
&&\left<\psi^{T}_{+}\epsilon \psi_{-}\right>=\left<\psi^{T}_{-}\epsilon \psi_{+}\right>= \Sigma\nonumber\\
&&\left<\bar \psi^{T}_{+}\epsilon \bar \psi_{-}\right>=\left<\bar \psi^{T}_{-}\epsilon \bar \psi_{+}\right>=\Sigma
\eea
that breaks the $Z_4$ symmetry of (\ref{NEWS}).
This is the mechanism by which the instanton-dyon liquid
enforces the anomalous $U_A(1)$ breaking with adjoint fermions.
The fermionic quadri-linears  in (\ref{NEWS}) can be reduced by introducing
pertinent Lagrange multipliers $\Lambda^\prime$s through the identity as detailed in~\cite{LIU2}. Assuming parity symmetry,
in  the mean-field or Hartree approximation,
(\ref{NEWS}) becomes

\bea
\label{SXX}
&&S\rightarrow S+\int d^3x\,\psi^{T} \tilde{\bf G}\psi \nonumber\\
&&+\sum_\pm \int d^3x\,\Lambda_1 (x)(\psi^{T}_{\pm}\epsilon \psi_{\mp}-\Sigma)\nonumber\\
&&+\sum_\pm \int d^3x \,{\Lambda}_2(x)(\bar \psi^{T}_{\pm}\epsilon \bar \psi_{\mp}-\Sigma)
\eea
 We observe that the mean-field constraints in (\ref{SXX}) enforce the substitution
 $\psi^{T}\epsilon\psi\rightarrow \Sigma $, and therefore the shift for $\Sigma\neq 0$

\be
e^{w_M-w_L}\rightarrow \sqrt{\frac{f_L}{f_M}}|\Sigma|e^{w_M-w_L}\nonumber\\
e^{w_L-w_M}\rightarrow \sqrt{\frac{f_M}{f_L}}\frac{1}{|\Sigma|}e^{w_L-w_M}
\ee
For completeness, 
the exchange or Fock correction to the mean-field approximation (\ref{HARTREE})  is detailed in Appendix D.
Also, a 1-loop alternative approximation is presented  in Appendix E.

To  insure a smooth limit for $\nu\rightarrow 1/2$, we will redefine the magnetic fugacity
$f_M(2\nu-1)^6\rightarrow  f_M$ throughout. As half the zero-modes jump when $\nu=1/2$, the hopping 
is singular in the ensemble made of constituent instanton-dyons and instanton-anti-dyons. This singularity does not appear
if the constituents are jumpng within the KvBLL caloron as all infrared tails are tamed, as we have shown in Appendix C.
But again, the fact that the delocalization of the zero-modes
makes use of the hopping between instanton-dyons and instanton-anti-dyons because of the
chirality flip, it is necessary to unlock the constituents from their respective KvBLL calorons
and anti-calorons as we have detailed.

\begin{figure}[h!]
  \begin{center}
  \includegraphics[width=6cm]{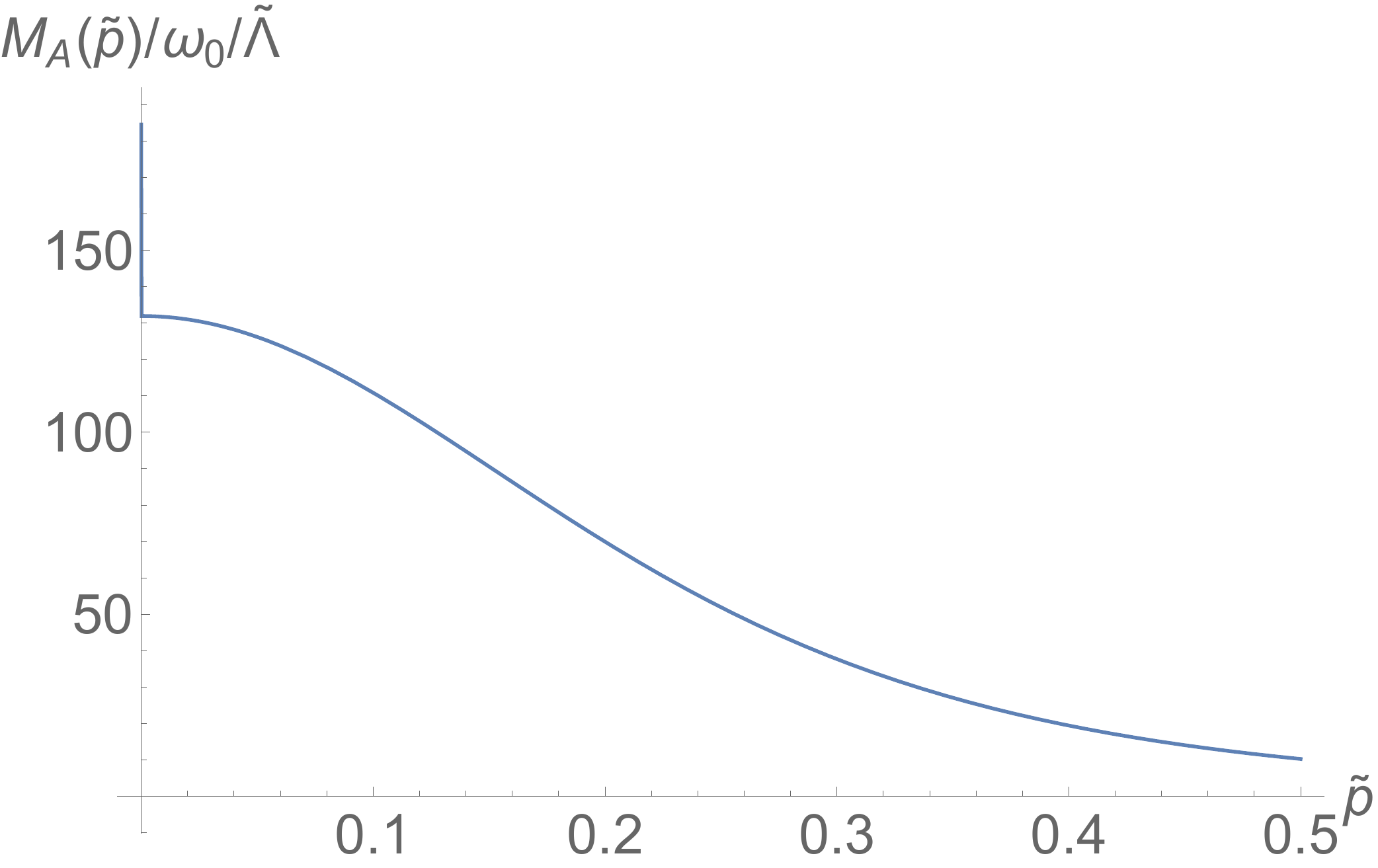}
   \caption{Adjoint constituent mass  for $\nu=0.7$.}
     \label{fig_map}
  \end{center}
\end{figure}

With the above in mind, a repeat of the quenched arguments show that the unquenched  pressure ${\cal P}_D=-{\cal V}/V_3$  with adjoint
and massless fermions is now

\bea
\label{PRESSUREX}
&&\frac{\cal P_{D+F}}{T^3}=-\frac{\tilde n_\Sigma^2}{8}\left(\frac{e^{2\bf b}}{\nu}+\frac{e^{-2\bf b}}{\bar \nu}\right)
+\tilde n_\Sigma\,(\nu e^{\bf  b}+\bar \nu e^{-\bf b})
\nonumber \\&&-4\tilde \Sigma \tilde \Lambda+\pi \int \tilde p^2d\tilde p\ln (1+\tilde \Lambda^2\mathbb F) +\frac{4\pi^2}3\nu^2\bar\nu^2\nonumber\\
\eea
with $\tilde n_\Sigma=8\pi f\Sigma/T^3$ and $\tilde\Lambda=\Lambda/T^2$.
We have defined

\bea
\label{FXX}
&&\pi^4\mathbb F(\tilde p,2\nu-1)=(3f_1^2+f_3^2+2f_1f_3-8f_2^2+8f_1f_2\tilde p)^2\nonumber \\
&&+ \tilde p^2\left(-f_1^2+f_3^2+2f_1f_3+8f_2^2+8f_1f_2\frac{1}{\tilde p}\right)^2
\eea
The $f_i$ are  given in (\ref{FFF}) after replacing $p\rightarrow \tilde p=p/\omega_0$ and $\tilde\nu\rightarrow \tilde\nu/\omega_0$,
all  of which are now dimensionless.  We have numerically checked that the momentum integration in
(\ref{PRESSUREX}) does not change much if we were to simplify the $f_i$ in (\ref{FFF}) to

\bea
&&f_1\approx -\frac{f_3}{3}\rightarrow  -\frac{1}{\tilde p^3}\tan^{-1}\left(\frac{\tilde p}{2\nu-1}\right)\nonumber\\
&&f_2\rightarrow \frac{\tilde p}{(\tilde p^2+(2\nu-1)^2)^2}
\eea
so that

\bea
\label{FXXX}
\mathbb F (\tilde p,2\nu-1)\approx &&\frac 1{\pi^4} (6f_1^2-8f_2^2-8f_1f_2\tilde p)^2 \nonumber\\
&&+\tilde p^2(2f_1^2+8f_2^2-8f_1f_2\frac{1}{\tilde p})^2
\eea
The integral contribution in (\ref{PRESSUREX}) is that of a constituent adjoint quark, with a momentum dependent
mass $M_A(\tilde p)$ given by

\be
\frac{M_A(\tilde p)}{\omega_0\tilde\Lambda}=\left((1+\tilde p^2)\,{\mathbb F}\right)^{\frac 12}
\ee
as shown in~Fig.~\ref{fig_map} for $\nu=0.7$.

\subsection{Confining symmetric phase}

The center-symmetric state with ${\bf b}=0$ and $\nu=1/2$ is an extremum of (\ref{PRESSUREX}),
provided that $\Sigma\neq 0$. This means that the spontaneous breaking of chiral symmetry is a
necessary (but not sufficient) condition for center-symmetry to take place in the instanton-dyon liquid model with massless adjoint
quarks. This  is similar to the observation made in~\cite{LIU2} for massless fundamental quarks.
For fixed $\tilde\Lambda $, the fermionic contribution in (\ref{PRESSUREX}) is maximal for $\nu=1/2$.
The additional extremum with respect to $\Sigma$ yields the  condition

\be
\label{XCX}
4\tilde \Lambda \tilde \Sigma =\tilde n_\Sigma(\nu e^{\bf b}+\bar \nu e^{-\bf b})
-\frac{\tilde n_\Sigma^2}{4}\left(\frac{e^{\bf b}}{\nu}+\frac{e^{-\bf b}}{\bar \nu}\right)
\ee
with  $\tilde n_\Sigma={n_\Sigma}/{T^3}$.
(\ref{XCX}) requires $\tilde n_\Sigma<1$  so that $\tilde\Lambda\neq 0$
and therefore a final quark condensate. We recall that for $N_f=0$, $\tilde n_\Sigma>\tilde n_D=0.56$ is required for
a center symmetric state. With this in mind, and for $0.56<\tilde n_\Sigma<1$, the extremum in the $\tilde\Lambda$ direction gives
the gap equation

\be
\label{XGAPX}
{\tilde n_\Sigma-\tilde n_\Sigma^2}=2\pi\int \tilde p^2d\tilde p\frac{\tilde \Lambda^2 \mathbb F }{1+\tilde \Lambda^2\mathbb F}
\ee
(\ref{XGAPX}) yields a finite $\tilde\Lambda$ and thus a finite chiral condensate. We note that a  core strength $V_0\rightarrow 0$
amounts to a vanishingly small $\tilde n_\Sigma^2\rightarrow 0$ contribution. Note that in the
center symmetric phase phase with $\tilde n_D\approx 1/2$, the core correction is about 50\% of the free instanton-dyon contribution.
It decreases substantially in the center asymmetric phase as the instanton-dyon liquid dilutes.

More explicitlly, for small $\tilde \Lambda$ the  dominant contributions from the hopping fermions stem from
the small momentum sector of the p-integrals in (\ref{PRESSUREX}) and (\ref{XGAPX}) with

\bea
\label{SMALLX}
\mathbb F (p\rightarrow 0, 0)\approx&&\frac{0.47}{\tilde p^{12}}
\eea
Inserting (\ref{SMALLX}) into (\ref{XGAPX}) allows for an explicit solution to the gap equation
in the form

\bea
\label{CXXX}
 \tilde \Lambda\approx \left(\frac{\tilde n_\Sigma-\tilde n_\Sigma^2}{1.92}\right)^2
\eea

\subsection{ The magnitude of the chiral condensate
}
For massive adjoint quarks, the fermionic part of  (\ref{PRESSUREX}) is 

\bea
{\pi}\int \tilde p^2d\tilde p
\,\ln \left((1+{\tilde m{\bf t}\tilde \Lambda)^2+{\tilde \Lambda}^2{\mathbb F}}\right)
\eea
where all contributions are dimensionless. We have defined

\bea
\label{TKX}
{\bf t}(p)=&&\frac{\omega _0^3}{\pi^{2}}\,{\bf K}(p)\nonumber\\
\tilde m=&&\frac{m}{\omega_0}
\eea
The chiral condensate for massless adjoint fermions follows from the general  relation

\bea
\label{CONYY}
&&\left<i{\rm Tr}( \lambda \lambda)\right>=\frac{1}{TV_3}
\left(\frac{\partial \ln Z}{\partial m}\right)_{m=0}\nonumber \\
&&=T^3 \int \tilde p^2 d\tilde p\frac{{2{\bf t}\,\tilde \Lambda}}{1+\tilde \Lambda^2\mathbb F}
\eea

Again, the integration in (\ref{CONYY}) is dominated by small momenta for small $\tilde\Lambda$.
In the confined state with $\nu=1/2$, we can use (\ref{SMALLX}) and the small momentum
limit of (\ref{TKX})

\be
{\bf t}(p\rightarrow 0)\approx \frac{2.31}{\tilde p^6}
\ee
to obtain

\be
\label{CONZZ}
\frac{\left<i{\rm Tr}( \lambda \lambda)\right>}{T^3}\approx 2\,\sqrt{\tilde \Lambda}\approx \, (\tilde n_\Sigma-\tilde n_\Sigma^2)
\ee
Again we note that for a vanishingly small core with $V_0\rightarrow 0$, the contribution $n_\Sigma^2\rightarrow 0$
in (\ref{CXXX}) with a chiral condensate for adjoint fermions of order $\tilde{n}$ which is the rescaled
instanton dyon density. This  result is totally consistent with the result derived in~\cite{LIU2} for massless
fundamental  quarks with no core.

The transition from a symmetric state with $\nu=1/2$ to an asymmetric state with
$\nu<1/2$ takes place $n_\Sigma<n_D$ as the instanton-dyon liquid dilutes, and the chiral condensate (\ref{CONZZ})  also vanishes
(see below).

\subsection{General case with $N_f\geq 1$}

The preceding analysis generalizes to $N_c=2$ and $N_f\geq 1$ adjoint fermions
through the substitution

\be
\psi^{T}_{+}\epsilon \psi_{-}\rightarrow \frac{1}{N_f!}\det_{fg}\psi^{T}_{+f}\epsilon\psi_{-g}
\ee
in (\ref{NEWS}) with all other labels unchanged. As a result the fermionic terms are  $SU(N_f)\times Z_{4N_f}$ flavor symmetric.
The $U_A(1)$ symmetry for adjoint QCD is explicitly broken by the instanton-dyon liquid model. The flavor symmetry is further broken spontaneously through $SU(N_f)\times Z_{4N_f}\rightarrow O(N_f)$ with the appearance of a condensate

\be
\label{CONDXXX}
\left<\psi^{T}_{+f}\epsilon \psi_{-g}\right>=\Sigma\delta_{fg}
\ee
the dual of the chiral condensate. (\ref{CONDXXX}) is explicitly symmetric under the transformations
$\psi_{\pm f}\rightarrow O_{fg}\psi_{\pm g}$ and $\bar \psi_{\pm f}\rightarrow \bar \psi_{\pm g}O^{T}_{gf}$.

A rerun of the preceding arguments yield the instanton-dyon plus adjoint fermions
pressure for arbitrary $N_f$

\bea
\label{PRESSUREXNF}
&&{\cal P}_{\cal D+ F}=-\frac{8\pi^2f^2\Sigma^{2N_f}}{T^3}\left(\frac{e^{2\bf b}}{\nu}+\frac{e^{-2\bf b}}{\bar \nu}\right)\nonumber \\
&&+8\pi f\Sigma^{N_f}(\nu e^{\bf b}+\bar \nu e^{-\bf  b})-4N_f\Lambda\Sigma\nonumber \\
&&+N_f\int \frac{d^3p}{(2\pi)^3}\ln(1+\Lambda ^2\tilde{\bf T}^{+2})+P_{\rm loop}(N_f)\nonumber\\
\eea
The last contribution is  briefly detailed in Appendix F and is seen to be dominated by the first term in the expansion. 
If we were to define $\tilde n_{\Sigma f}=8\pi f\Sigma^{N_f}/T^3$ then  the results from (\ref{PRESSUREXNF})
for arbitrary $N_f$  map onto  those from
(\ref{PRESSUREX})
for $N_f=1$,  with now

\bea
\label{PRESSUREZ}
&&\frac{{\cal P_{D+F}}}{T^3}
=-\frac{\tilde n_{\Sigma f}^2}{8}\left(\frac{e^{2\bf b}}{\nu}+\frac{e^{-2\bf b}}{\bar \nu}\right)+\tilde n_{\Sigma f}\,(\nu e^{\bf  b}+\bar \nu e^{-\bf b})
\nonumber \\&&-4N_f\tilde \Sigma \tilde \Lambda+\pi N_f\int \tilde p^2d\tilde p\ln (1+\tilde \Lambda^2\mathbb F) \nonumber\\
&&-\frac{4\pi^2}{3}(1+N_f)\nu^2\bar \nu^2
\eea
The ground state is center symmetric for a sufficiently dense instanton-dyon liquid, provided that chiral symmetry
is spontaneously broken with $\Sigma\neq 0$, and symmetric  in the dilute limit. Here $\tilde\Lambda, \tilde\Sigma$ 
follow from the extrema of (\ref{PRESSUREZ}) as coupled gap equations,

\bea
\label{GAPXX}
&&\tilde\Sigma = \frac {\pi}2\int\,\tilde p^2d\tilde p\frac{\tilde \Lambda\mathbb F}{1+\tilde \Lambda^2\mathbb F}\nonumber\\
&&\tilde\Lambda\tilde\Sigma=
-\frac{\tilde n_{\Sigma f}^2}{16}\left(\frac{e^{2\bf b}}{\nu}
+\frac{e^{-2\bf b}}{\bar \nu}\right)+\frac{\tilde n_{\Sigma f}}4 (\nu e^{\bf  b}+\bar \nu e^{-\bf b})\nonumber\\
\eea
The solutions $\tilde\Sigma({\bf b}, \nu)$ 
and $\tilde\Lambda({\bf b}, \nu)$ to (\ref{GAPXX}) should be inserted back 
 in (\ref{PRESSUREZ}) to maximize numerically  the pressure in 
the parameter space $\nu, {\bf b}$. 

\begin{figure}[h!]
  \begin{center}
  \includegraphics[width=8cm]{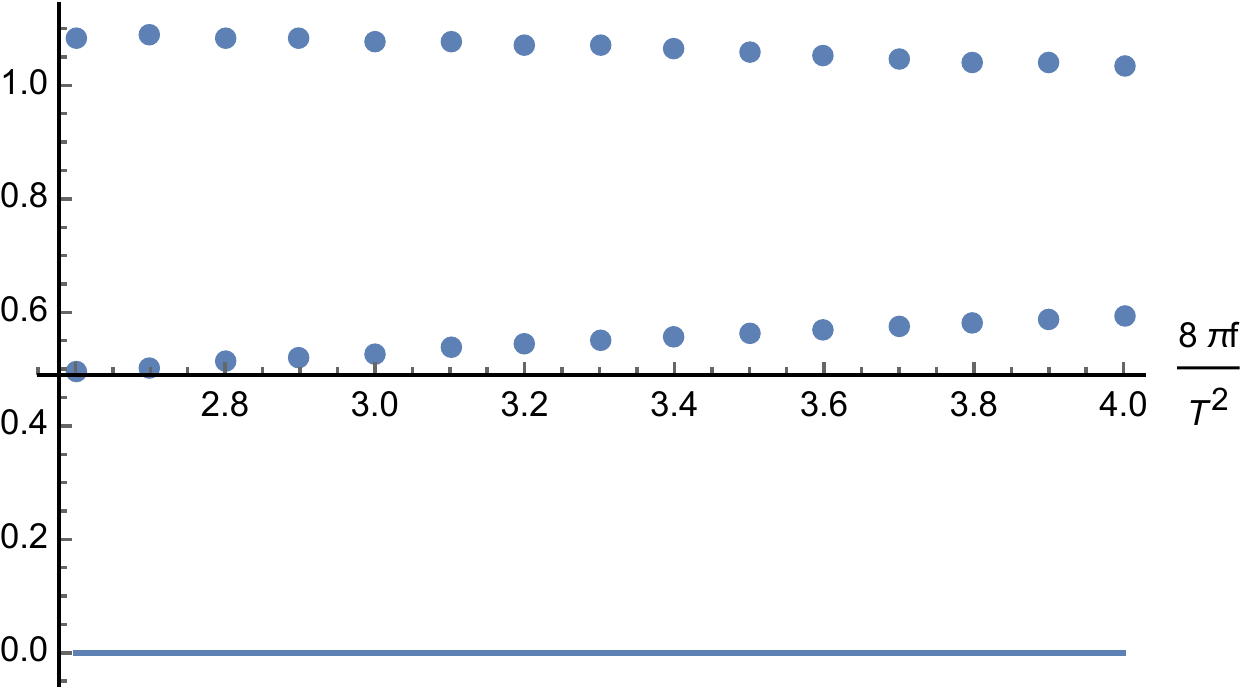}
   \caption{Dimensionless pressure (middle dotted line), Polyakov line (solid line) and chiral condensate (upper dotted line) 
   versus $8\pi f/T^2$ (decreasing temperature) for $N_f=1$.}
     \label{fig_PLC}
  \end{center}
\end{figure}

In Fig.~\ref{fig_PLC} we show the numerical results for 
the dimensionless pressure (dotted middle line), Polyakov line (solid line) and chiral condensate (dotted upper line)
with increasing $8\pi f/T^2$ (decreasing temperature), for $N_f=1$ in the symmetric phase. The spontaneous breaking of
chiral symmetry is lost for $8\pi f/T^2<2. 6$, which causes all topological effects to vabish in the chiral limit.
For $N_f>2$,   (\ref{PRESSUREZ}-\ref{GAPXX}) do not support a solution that breaks spontaneously chiral symmetry.

Finally, the restoration of chiral symmetry
 can be estimated analytically from (\ref{PRESSUREZ}-\ref{GAPXX}), by dropping the first or core contribution, and noting that the resulting 
expression maps onto the  one derived for fundamental quarks in~\cite{LIU2} (see Eq. (80) there) with $N_cN_f$.
This mapping shows that (\ref{PRESSUREZ}-\ref{GAPXX}) does not sustain a chiral condensate for $N_cN_f/N_c\geq 2$, or
$N_f\geq 2$ Majorana quarks.

\subsection{Critical temperature estimates}

For general $N_f$,  we can estimate the critical temperature for the restoration of center symmetry $T_D$,
by neglecting both the core and fermionic contributions in (\ref{PRESSUREZ}), i.e.

\bea
\frac{{\cal P_{D+F}}}{T^3}
\rightarrow 
\tilde n_{\Sigma f}\,(\nu e^{\bf  b}+\bar \nu e^{-\bf b})-\frac{4\pi^2}{3}(1+N_f)\nu^2\bar \nu^2
\eea
An estimate of the deconfining temperature $T_D$ follows by balancing the first contribution
in the center symmetric phase with ${\bf b}=0$ and $\nu=\bar \nu=1/2$, against the last 1-loop
contribution  stemming from the adjoint free gluons and quarks. The result is

\be
\label{TDXX}
\frac{ n_{\Sigma f} }{T_D^3}\approx \frac {\pi^2}{12}(1+N_f)
\ee
In the presence of adjoint quarks, the  fundamental string tension does not vanish, 
$\sigma/T^2= n_{\Sigma f}/T^3$. For $N_c=2$  QCD with $N_f$ adjoint Majorana quarks, the 
ratio of the critical  temperature for center symmetry loss normalized by the fundamental
string tension decreases with $N_f$ as

\be
\label{TDXXX}
\frac{T_D}{\sqrt{\sigma}}\approx \frac 2\pi \left(\frac 3{1+N_f}\right)^{\frac 12}
\ee
It would be useful to check (\ref{TDXXX}) against current lattice simulations
with adjoint quarks. 

The estimate of the chiral symmetry restoration temperature for the chirally broken phase with $N_f<2$, is more
subtle. For that we recall, that the delocalization of the adjoint zero modes generates
the so-called zero-mode-zone with a finite eigenvalue density  $\rho(\lambda)$ 
normalized to the 4-volume $V_3/T$. 
The details of the interactions in the small virtuality $\lambda$ limit do not matter~\cite{RMTX}, as the
distribution follows Wigner semi-circle

\be
\label{DENSX}
\rho(\lambda)=\frac{4n_{\Sigma f}}{(\lambda_{\rm max}(T)/T)}\left(1-\frac {\lambda^2}{\lambda_{\rm max}^2(T)}\right)^{\frac 12}
\ee
The normalization is fixed by the overall number of zero modes in the instanton-dyon liquid.
Here $2\lambda_{\rm max}(T)$ is the size of the zero-mode-zone at finite $T$.
Combining (\ref{CONZZ}) with the Banks-Casher relation~\cite{BK} we have

\be
|\tilde n_{\Sigma f}-\tilde n_{\Sigma f}^2|\approx\pi\rho(0)
\ee
which fixes $x(T)=\lambda_{\rm max}(T)/(\pi T)$ as

\be
\label{TCXX}
{ \tilde n_{\Sigma f} }\approx 1-\frac{2}{x(T)}
\ee
The chiral transition temperature $T_C$ is fixed by the quarks turning massless
or $\Sigma\rightarrow 0$ which implies that the  instanton-dyon density 
$\tilde n_{\Sigma f}\rightarrow 0$, as all topological contributions are suppressed. 
From (\ref{TCXX})  this occurs when

\be
\label{XTC}
T_C=\frac{\lambda_{{\rm max}}(T_C)}{2\pi}
\ee

We now note that  at the chiral transition temperature, the quark hopping stalls 
into topologically neutral molecules. As a result $\tilde{\bf T}$ in 
(\ref{SU2}) becomes banded, and $\lambda_{+}(T_C)$ is comparable to the strength of the nearest
neighbor hopping (\ref{T++})

\bea
\label{XTCC}
\lambda_{\rm max}(T_C)=&& \left|{\bf T}^+(x_{IJ}=0)\right|\nonumber\\
= &&\left|\int \frac{d^3p}{(2\pi)^3}\, 
{\bf T}^+(p)\right|=\kappa\,\pi T_C\,\left|2\nu_C- 1\right|\nonumber\\
\eea
with $\kappa=0.557$. Using  (\ref{XTC}-\ref{XTCC}), it follows that chiral restoration occurs 
when the holonomy reaches $\nu_C=1/2+1/\kappa=0.3$  (${\rm mod}\, 1$), and in general
$T_C>T_D$. 

Using the quenched effective potential discussed earlier for an estimate, this corresponds to an instanton-dyon density for chiral restoration $\tilde n_C=0.48$, which is  surprisingly close to the quenched instanton-dyon density for 
the breaking of center symmetry $\tilde n_D=0.56$.  Using the  instanton-dyon density
for $N_c=2$ and $N_f=1$ Majorana quark

\bea
&&\tilde n(T)\approx Ce^{-\pi/\alpha_s(T)}\approx C\left(\frac{0.36T_D}T\right)^{\frac {21}6}
\eea
we find that 

\be
\left(\frac{T_C}{T_D}\right)\approx \left(\frac{0.56}{0.48}\right)^{\frac 6{21}}\approx  1
\ee
which is much smaller than the ratio reported in  lattice simulations~\cite{LATTICEX}.

\section{Conclusions}

We have presented a mean-field analysis of key characteristics of the instanton-dyon liquid
with adjoint light quarks.  The index theorem on $S^1\times R^3$ shows that dissociated
instanton-dyons support 4 anti-periodic zero modes, that localize on the M-instanton-dyon in the
center asymmetric phase with  $\nu>1/2$, or alternatively, on the
L-instanton-dyon for $\nu<1/2$.  These two cases are dual to each other, so only one can be considered.
In the symmetric phase, the 4 anti-periodic zero modes are shared
equally (two on each) by the L- and M-instanton-dyons. We have used the ADHM construction to derive
explicit form of these zero modes.

We have detailed the construction of the partition function for the  dissociated KvBLL
calorons with $N_f$ light adjoint quarks, including the  classical  streamline interactions and
the quantum Coulomb interactions induced by the coset manifold. We have retained a core
interaction between the like instanton-dyon-antidyons to distinguish them from perturbative fluctuations.
By a series of fermionization
and bosonization techniques, we have mapped this interacting many-body system on a 3-dimensional
effective theory. We have presented a mean-field analysis of the dense phase, that exhibits both  confinement
with center symmetry, and spontaneously broken chiral symmetry. We have shown that 
in such an approximation the deconfinement with  breaking
of center symmetry, and  the restoration of chiral symmetry occur about simultaneously. Furthermore, the latter is
always unbroken for $N_f\geq 2$.

The mean-field analysis we have presented has a major shortcoming as the instanton-dyon liquid
dilutes. It does not account for the molecular pairing of the instanton-dyon-anti-dyon configurations
through light adjoint pairs.
We have presented 
a qualitative argument for the chiral transition using the assumption of pairing, but a more reliable
analysis is likely numerical as the analysis  goes beyond the mean-field results presented here.

\section{Acknowledgements}

This work was supported by the U.S. Department of Energy under Contract No.
DE-FG-88ER40388.

\section{Appendix A: Periodic zero-modes}

In this Appendix we briefly detail the ADHM construct  as applied to the periodic adjoint zero modes.
This is partly a check on our general ADHM construction.
For that we note that the Grassmanian matrix for periodic adjoint fermions simplifies to

\be
M(z,z^\prime)=\delta (z-z^\prime)M
\ee
A rerun of the preceding arguments yields the periodic zero modes

\bea
&&\lambda_{m}(r)=\frac 1{\rm {sh (\omega_0 r)}}\nonumber\\
&&\times \left(a (\omega_0 r)\sigma_m+b( \omega_0 r)\sigma \cdot  \hat r\sigma _m \sigma\cdot \hat r)\epsilon M\right.\nonumber \\
&&\left.- \epsilon (M^Ta(\omega_0 r)\sigma_m+M^Tb(\omega_0 r)\sigma \cdot  \hat r \sigma _m \sigma\cdot \hat r)^T\right)\nonumber\\
\eea
For $\omega_0r\rightarrow\infty$, we have
$a\approx b\approx  -{{\rm sinh}(\omega_0 r)}/{(\omega_0r)^2}$, so that

\bea
\label{ZEROPER}
\lambda_m (r\rightarrow\infty) =&& \frac{1}{r^2}(\sigma_m+\sigma \cdot \hat r\sigma _m\sigma \cdot \hat r)
\chi\nonumber\\ =&& \frac{2}{r^2}r_m \sigma \cdot \hat r \chi
\eea
with $\chi=\epsilon M$. (\ref{ZEROPER}) are in agreement with the known periodic zero modes in
the hedgehog gauge~\cite{MATTIS,DPX}.

\section{Appendix B:  Zero-modes in a BPS Dyon without ADHM}

In this Appendix we  explicitly derive the Dirac equation for anti-periodic adjoint fermions in the
state of lowest total angular momentum, without using the ADHM construction. We will use the
equations to investigate the nature of the fermionic zero mode at the origin and asymptotically.
Without the ADHM construct, the equations are only solvable numerically.

Without loss of
generality, we will consider the $\overline M$-dyon gauge configuration given by

\bea
\label{MBAR}
\left(A_4^a, A^a_{i}\right)=\left({\hat r}^a\phi(r), \epsilon_{aij}{\hat r}_jA(r)\right)
\eea
with the boundary values

\bea
&&A(r\rightarrow 0)=0\qquad A(r\rightarrow \infty)=-\frac{1}{r}\nonumber\\
&&\phi(r\rightarrow 0)=0\qquad \phi(r\rightarrow \infty)=2\pi T\nu
\eea
In the adjoint representation of $SU_c(2)$ the color matrices are
$T^a_{mn}=i\epsilon_{amn}$. In the chiral basis, the adjoint Dirac fermions
will be sought in the form

\be
\label{ISO}
\Psi\equiv
\begin{pmatrix}
\Psi^+_m\\
\Psi^-_m\\
\end{pmatrix}
\ee
The Dirac equation (\ref{DIR}) for the 2 lowest Matsubara frequencies $\pm\omega_0$
is given by

\bea
\label{ADJ1}
&&\left(i\sigma \cdot \nabla \delta_{nm}+i(\sigma_{n}\hat r_{m}-\sigma_{m}\hat r_n){A(r)}\right.\nonumber\\
&&\left.\pm\epsilon_{nam}\hat r_a{\phi(r)}\right)\Psi_m^{\pm}=i\omega_0\Psi_n^{\pm}
\eea

To solve (\ref{ADJ1}) explicitly, we
decompose the vector-valued chiral components in (\ref{ISO}) using the independent vector
basis~\cite{REBBI}

\be
(1, \vec\sigma \cdot \hat r)\, \left(\hat r, (\vec r\times \vec p), (\vec r\times \vec p)\times \hat r\right)
\ee
which is seen to commute with the total angular momentum $\vec J=\vec l+\vec s$.
We seek the zero modes in the state of zero orbital angular momentum or $J=1/2$. Therefore,

\bea
\label{ADJ2}
&&\Psi^{\pm}_m(\vec r)\equiv \hat r_m\,\Theta^{\pm}_{3}+(\vec r\times\vec p)_m\,(\sigma\cdot \hat r)\,\Theta^{\pm}_4\nonumber \\
&&+\hat r_m\,\sigma\cdot \hat r \,\Theta^{\pm}_{1}+i((\vec r\times \vec p)\times \hat r)_m(\sigma \cdot \hat r )\,\Theta^{\pm}_2
\eea
with the scalar radial spinor-functions

\be
\label{ADJ3}
\Theta^{\pm}_{i}\equiv \sum_{s=\pm } F^{\pm}_i(r,s)\left.|s\right>
\ee
Inserting (\ref{ADJ2}-\ref{ADJ3}) into (\ref{ADJ1}) yield

\bea
&&\left(\frac{d}{dr}+\frac{2}{r}\right)F^{\pm}_1-2\rho F_2^{\pm}=\omega_0F^{\pm}_3\nonumber\\
&&\left(\frac{d}{dr}+\frac{1}{r}\pm\phi\right)F^{\pm}_2-\rho F^{\pm}_{1}=\omega_0F^{\pm}_4\nonumber\\
&&\frac{d}{dr}F^{\pm}_3+2\rho F^{\pm}_4=\omega_0F^{\pm}_1\nonumber\\
&&\left(\frac{d}{dr}+\frac{1}{r}\pm\phi\right)F^{\pm}_4+\rho F^{\pm}_3=\omega_0 F^{\pm}_2
\eea
Here $\rho\equiv\left<A_4\right>+{1}/{r}$, with the label $s$ subsumed.
Using the asymptotics, it is readily found at infinity that

\bea
\label{LZ}
&&F^{\pm}_{1,3}(r\rightarrow\infty) = c_1e^{-\omega_0r}+c_2e^{+\omega_0 r}\nonumber\\
&&F^{\pm}_{2,4}(r\rightarrow\infty) = c_3 e^{-\omega_0 (1\pm 2\nu)r}+c_4e^{+\omega_0(1\mp 2\nu) r}\nonumber\\
\eea
while at the origin we have

\bea
&&F^\pm_{3,4}(r\rightarrow 0)= b_3r+b_4\frac{1}{r^2}\rightarrow b_3r\nonumber\\
&&F^\pm_{1,2}(r\rightarrow 0) = b_1+b_2\frac{1}{r^3}+ b_3r^2+b_4\frac{1}{r}\rightarrow b_1+b_3r^2\nonumber\\
\eea
For $F^+$ with fixed $s=\pm$,
we always have 2 ($b_{1,3}$) out of 4 ($b_{1,2,3,4}$) total dimension of solutions which are normalizable at 0.
We have 2 ($c_{1,3}$) out of 4 ($c_{1,2,3,4}$) total dimension of solutions which are normalizable
at infinity for $\nu\le\frac{1}{2}$, and 3 ($c_{1,3,4}$) for $\nu>1/2$ . We conclude that for
$\nu>\frac{1}{2}$, there exists at least one zero mode. For $\nu<\frac{1}{2}$, the existence cannot be proved
on general grounds and a numerical analysis is required. However, their existence is supported by the index theorem
reviewed earlier.  For $\nu>1/2$, the dominant contribution at large distances stem from the asymptotic in
(\ref{LZ}) or $c_4e^{-(2\nu-1)\omega_0r}$.  As $\nu\rightarrow 1/2$, it asymptotes a constant which is not square
integrable. This analysis for $\nu=1/2$ requires more care, as we discussed earlier in the ADHM construction.

\section{Appendix C: Adjoint fermions in a KvBLL Caloron}

The adjoint fermions in the classical background of KvBLL calorons  can be constructed
using the general ADHM construct presented above. For an alternative derivation using 
the replica trick for adjoint fermions in calorons we refer to~\cite{ADHMX}. 
We recall that the BPS dyon results
follow  by taking various limits.  The matrix of ADHM data is more involved in a KvBLL caloron.
For the $SU(2)$ KvBLL caloron with a holonomy $P_{\infty}=e^{i2\pi \omega \cdot \sigma}$
and $\omega=\nu/2\beta\rightarrow  \nu/2$, we have for the quaternionic blocks

\bea
\lambda(z)=&&(P_{+}\delta(z-\omega)+P_-\delta(z+\omega))q\nonumber\\
B(z,z^\prime)=&&\delta(z-z^{\prime})\left(\frac{1}{2\pi i}\frac{\partial }{\partial z^\prime}+A(z^\prime)\right)
\eea
with $P_\pm$ as projectors and

\be
A(z)=\chi_{[-\omega,\omega]}(z)+\bar q \omega \cdot \sigma q\,(\chi_{[-\omega,\omega]}(z)-2\omega)
\ee

The periodicity of the gauge field $A_m(x_4+\beta)=A_m(x_4)$  (modulo a gauge transformation)
and the anti-periodicity of the adjoint fermions yield

\bea
c_m=&&-e^{2\pi i\omega \cdot \sigma}c_{m-1}\nonumber\\
\bar c_m=&&-\bar c_{m}e^{-2\pi i\omega \cdot \sigma}\nonumber\\
M_{mn}=&&-M_{m-1,n-1}
\eea
Their Fourier transforms are

\bea
\label{APP1}
c(z)=&&\left(P_+\delta\left(z-\omega+\frac 12\right)\right.\nonumber\\
&&\left.+P_-\delta\left(z+\omega+\frac 12\right)\right)\,c\nonumber\\
\bar c(z)=&&\bar c\left(P_+\delta\left(-z-\omega+\frac 12 \right)\right.\nonumber\\
&&\left.+P_-\delta\left(-z+\omega+\frac 12\right)\right)\nonumber\\
M(z,z^\prime)=&&\delta\left(z-z^\prime+\frac 12\right)\,M(z^\prime)
\eea
Inserting (\ref{APP1}) in the adjoint zero mode constraint gives

\bea
&&\frac{1}{2\pi i}\frac{d}{dz}M(z)+\left(A^T(z)-A^T\left(z+\frac 12\right)\right)\,M(z)\nonumber \\
&&-\epsilon_2\, \bar qP_{+}c\, \, \delta\left(z+\omega+\frac 12\right)
-\epsilon_2\, \bar q P_-c\,\, \delta\left(z-\omega+\frac 12\right)\nonumber \\
&&-q^{T}P_+^T\bar c^T\,\,\delta(z+\omega)-q^{T}P_-^T\bar c^T\,\,\delta(z-\omega)=0\nonumber\\
\eea
The explicit form of the zero modes are

\bea
&&(\lambda_{\alpha})_{ab}\,\phi(x)=\int^{+\frac 12}_{-\frac 12} dz dz^\prime\\
&&\times((-c_a(-z)+u^{\dagger}_{a \beta }(z+1/2)(\epsilon M)_\beta(z))f(z,z^\prime)u_{\alpha b}(z^\prime)\nonumber \\
&&-u^{\dagger}_{ a\beta}(z)\epsilon _{\alpha \beta}f(z,z^\prime)(-\bar c_b(z^\prime+1/2)+ M_{\gamma}(z^\prime) u_{\gamma b}(z^\prime )))\nonumber
\eea
\bea
&&\lambda_m\,\phi(x) =\int^{+\frac 12}_{-\frac 12} dz dz^\prime\\
&&\times((u(z^\prime)f^{\star}(z^\prime ,z)\sigma_m(-c(-z)+u^\dagger(z+1/2)(\epsilon M)(z))\nonumber \\
&&-\epsilon((-\bar c(z^\prime+1/2)+M^T(z^\prime)u(z^\prime))\sigma_mf(z,z^\prime)u^{\dagger}(z))^T)\nonumber
\eea
with $\phi(x)=1+u^\dagger(x)u(x)$. Here the m-summation and z-integration are subsumed. The
x-argument in $u(x,z)$ has been omitted for convenience.

\subsection{Special case $\nu=\frac 12$}

For the center symmetric case with $\omega=1/2\nu=1/4$, we set
$\omega \cdot \sigma=\tau_3/4$ and $q=i\rho \tau_3$, and identify the
coordinates of the constituents $M,L$ of the KvBLL caloron as

\bea
{\bf r}=&&x\cdot \sigma+\pi \rho^2 \tau_3/2\nonumber\\
{\bf s}=&&x\cdot \sigma -\pi \rho^2\tau_3/2
\eea
in terms of which

\be
A(z)-x=-i{\bf s}\chi_{[-1/4,1/4]}(z)-i{\bf  r}\chi_{[1/4,3/4]}(z)\equiv -i{\bf R}(z)\nonumber\\
\ee
In this case, the equation for  $M$ simplifies

\bea
\epsilon M= e^{\pi \rho^2 \tau_3z}M_0 && -1/4<z<1/4\nonumber \\
\epsilon M=e^{-\pi \rho^2 (z-1/2)}M_0 && +1/4<z<3/4
\eea
and  $\bar c^T=-\epsilon c$.  The  C-zero-mode and M-zero-mode decouple,
with respectively

\bea
\lambda^C_m\phi(x)=&&\int_{-\frac 12}^{+\frac 12}  dzf\left(3/4 ,z\right)u(x,z)\sigma_mP_+c\nonumber \\
+&&\int_{-\frac 12}^{+\frac 12}  dz f\left(1/4 ,z\right)u(x,z)\sigma_mP_-c
\eea
and

\bea
\lambda^M_m\phi(x)=&&\int_{-\frac 12}^{+\frac 12}  f(z_1,z_2)u(x,z_2)\sigma_m u^{\dagger}(x,z_1+1/2)\nonumber \\
&&\times \epsilon (M(z_1)+M(-z_1-1/2))dz_1 dz_2
\eea
Here $u(z)$ is solution to the inhomogeneous and linear differential equation with piece-wise potential

\bea
&&\left(\frac{1}{2\pi i}\frac{\partial }{\partial z}+i{\bf R}(z)-x_4\right)u(x,z)\nonumber\\
&&=-i\tau_3 \rho(P_+\delta(-z+1/4)+P_-\delta(-z-1/4))
\eea
with the projectors $P_\pm =(1\pm \tau^3)/2$.
The explicit solutions are

\bea
u(x,z)=e^{2\pi i x_4 z}e^{2\pi {\bf s}z}B_1 (x) &&-1/4<z<1/4\nonumber\\
u(x,z)=e^{2\pi i x_4 (z-1/2)}e^{2\pi {\bf r}(z-1/2)}B_2 (x) &&+1/4<z<3/4\nonumber\\
\eea
and satisfy the completeness relations

\bea
&&e^{-\pi ix_4/2}e^{-\pi\bf{r}/4}B_2(x)-e^{\pi ix_4/2}e^{\pi\bf{s}/4}B_1(x)=+2\pi\rho P_+\nonumber\\
&&e^{-\pi ix_4/2}e^{-\pi \bf{s}/4}B_1(x)-e^{\pi ix_4 /2}e^{\pi \bf {r}/4}B_2(x)=-2\pi\rho P_-\nonumber\\
\eea
Here $B_{1,2}(x)$ are defined in Appendix C.
The solutions obey the quasi-periodicity conditions

\bea
u(x_4+1,{\bf x},z)=&&e^{2\pi iz}u(x_4,{\bf x},z)e^{-\pi \tau_3/2}\nonumber\\
B_1(x_4+1, {\bf x})=&&B_1(x_4, {\bf x})e^{-\pi \tau_3/2}\nonumber\\
B_2(x_4+1, {\bf x})=&&-B_2(x_4, {\bf x})e^{-\pi \tau_3/2}
\eea

With the above in mind, the explicit form of the C-zero mode is

\bea
\label{X0X}
&&\lambda^C_m\phi(x)=\\
&&(f_1+\hat s \cdot \sigma f_2)B_1\sigma_mP_+c+(\tilde f_1+\hat s\cdot \sigma \tilde f_2)B_1\sigma_mP_-c\nonumber \\
&&+(g_1+\hat r \cdot \sigma g_2)B_2 \sigma_mP_+c+(\tilde g_1+\hat r \cdot \sigma\tilde g_2)B_2\sigma_mP_-c\nonumber
\eea
where we have set  $s\equiv \omega_0 |\vec s|$ and $r=\omega_0 |\vec r|$. Also, we have

\bea
&&sr\psi (s,r,x_4)f_1(x_4, r,s)=\frac{e^{-\frac{1}{2} i \pi  x_4}}{4 s}(s+\sinh (s))\nonumber \\
&&\times \left(\sinh \left(\frac{s}{2}\right) \left(d \sinh (r)+r e^{2 i \pi  x_4}+r \cosh (r)\right)\right.\nonumber \\
&&\left.+s \sinh (r) \cosh \left(\frac{s}{2}\right)\right)
\eea
with $\psi$ given below, $d=\pi\rho^2$ and

\bea
&&sr \psi (s,r,x_4)f_2(x_4,r,s)=-\frac{e^{-\frac{1}{2} i \pi  x_4}}{4 s}(s-{\rm sinh}(s))\nonumber\\
&&\times
\left(-\cosh \left(\frac{s}{2}\right) \left(d \sinh (r)+r \left(-e^{2 i \pi  x_4}\right)+r \cosh (r)\right)\right.\nonumber \\
&&\left.-s \sinh (r) \sinh \left(\frac{s}{2}\right)\right)
\eea
with the following identities among the $f,\tilde f, g, \tilde g$ functions

\bea
&&\tilde f_1\equiv f_1(-x_4,{\bf x}),\qquad \tilde f_2\equiv-f_2(-x_4,{\bf x})\nonumber\\
&&g_1\equiv \tilde f_1(x_4,s,r),\qquad g_2\equiv \tilde f_2(x_4,s,r)\nonumber\\
&&\tilde g_1\equiv  g_1(-x_4,{\bf x}),\qquad \tilde g_2\equiv -g_2(-x_4,{\bf x})
\eea

\begin{figure}[h!]
  \begin{center}
  \includegraphics[width=10cm]{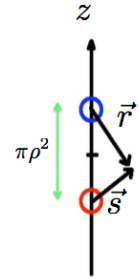}
   \caption{L-M-dyon at a distance $d=\pi\rho^2$ in a KvBLL caloron.}
     \label{fig_LM}
  \end{center}
\end{figure}

\subsection{Adjoint zero mode for Dyon from KvBLL caloron}

To isolate the adjoint zero modes on the constituents of the KvBLL caloron we take the limit
$d, |\vec r|\rightarrow\infty$ but fixed  $s$ fixed, which means that $r\rightarrow\infty$
as shown in Fig.~\ref{fig_LM}.   Most of the expressions simplify. Specifically, we have

\bea
\label{X1X}
&&f_1(x)=\frac{e^{-\frac{1}{2} i \pi  x_4}}{2s^2}
\frac{(s+\sinh (s)) \sinh(\frac{s}{2})}{({\rm cosh}(s)+\cos \theta {\rm sinh} (s))}\nonumber\\
&&f_2(x)=\frac{e^{-\frac{1}{2} i \pi  x_4}}{2s^2}
\frac{(s-\sinh (s))\cosh (\frac{s}{2})}{(\cosh(s)+\cos \theta \sinh (s))}
\eea
with $s\equiv \omega_0|\vec s|$, $\cos \theta=\vec s\cdot \hat z$, and

\bea
B_1=&&4\pi \rho(-\cos (\pi x_4))\nonumber\\
&&\times\left(\cosh \left(\frac{s}{2}\right)\tau_3+\sinh\left(\frac{s}{2}\right)\hat s\right)\nonumber \\
&&+i e^{\pi ix_4 \tau_3-\frac{i\pi }{2}\tau_3 x_4}\sin(\pi x_4)\nonumber\\
&&\times  \frac{(\cosh(\frac{s}{2})+\sinh(\frac{s}{2})\hat s \tau_3))}{\cosh (s)+\cos \theta \sinh (s)}\nonumber\\
B_2\rightarrow &&0
\eea
with also

\bea
\label{X2X}
\psi=&&e^{r}(\cosh(s)+\cos \theta \sinh (s))\nonumber\\
\phi=&&\frac{2d \cosh (s)}{s(\cosh(s)+\cos \theta \sinh (s))}
\eea
Inserting (\ref{X1X}-\ref{X2X}) into (\ref{X0X}) yields
the asymptotic zero mode on the localized instanton-dyon

\bea
\label{XEQ4}
&&s\cosh(s)(\cosh(s)+\cos \theta \sinh (s))\lambda^C_m \nonumber \\
&&=e^{-\frac{i\pi x_4}{2}}(sB_+ +\sinh(s)B_-)e^{-\frac{\pi i \tau_3x_4}{2}}B\sigma_mP_+c\nonumber\\
&&+e^{\frac{i\pi x_4}{2}}(sB_- +\sinh(s)B_+)e^{-\frac{\pi i \tau_3x_4}{2}}B\sigma_mP_+c
\eea
with

\bea
B_{\pm}=&&\sinh\left(\frac{s}{2}\right)\pm \hat s\cdot \sigma\cosh \left(\frac{s}{2}\right)\nonumber\\
B=&&\cosh\left(\frac{s}{2}\right)\tau_3+\sinh\left(\frac{s}{2}\right)\hat s\cdot \sigma
\eea

\subsection{String gauge}

The dyon reduced zero-mode from the KvBLL caloron (\ref{XEQ4})
carries a $\theta$-dependence contrary to  (\ref{EQ4}). (\ref{XEQ4}) is
expressed in the quasi-string gauge, while (\ref{EQ4}) is in the hedgehog gauge.
To express (\ref{XEQ4}) in the string gauge, we first gauge transform it
using $g=e^{i2\pi \omega \cdot \tau}$, to obtain

\bea
&&s \sinh(s)(\cosh(s)+\cos \theta \sinh(s))\lambda_b=\nonumber \\
&&e^{-i\omega_0 x_4}(P_+c)_a(sB_+B+\sinh(s)B_-B)_{\alpha b}\nonumber \\
&&+e^{i\omega_0 x_4}(P_-c)_a(sB_-B+\sinh(s)B_+B)_{\alpha b}
\eea
In the same gauge, the dyon gauge field reads

\bea
A_4=&&\tau_3\partial_3\ln \kappa+\kappa\tau_{\perp}\cdot \partial_{\perp}\zeta+2\omega \tau_3\nonumber\\
A_i=&&\tau_3\epsilon _{ij3}\partial_3 \ln \kappa+\kappa\tau_\perp \cdot \epsilon_{\perp ij}\partial_j \zeta\nonumber \\
&&+4\pi \omega\theta \kappa(\delta_{i1}\tau_2-\delta_{i2}\tau_1)
\eea
with

\bea
\zeta=&&\frac{4\pi \omega r}{\sinh(4\pi \omega r)}\nonumber\\
\zeta\kappa=&&\frac{1}{\cosh(4\pi \omega r)+\cos(\theta)\sinh(4\pi \omega r)}
\eea
which is still not in the string gauge. To bring the configuration (\ref{GG}) to the string gauge,
we make use of

\be
{\bf U}=\frac{\cosh(s/2)\tau_3+\sinh(s/2)\sigma \cdot s}{\sqrt{\cosh(s)+\cos(\theta)\sinh(s)}}
\ee
which is unitary.

\subsection{Definitions}

The matrices $B_{1,2}$ and the function $\psi$
are in agreement with those used in~\cite{DPX}. We quote them here for completeness.
Specifically

\bea
B_1=&&b_{12}b_{11}\,e^{-i2\pi x+4\omega\tau_3}\mathbb U^\dagger/\psi\nonumber\\
B_1=&&b_{22}b_{21}\,e^{-i2\pi x+4\omega\tau_3}\mathbb U^\dagger/\psi
\eea
with $\mathbb U$ a unitary color rotation and

\bea
b_{11}=&&i2\pi\rho\left(\overline{\cosh_{\frac 12}}+\hat r\tau_3 \overline{\sinh_{\frac 12}}\right)e^{i\pi x_4\tau_3}\nonumber\\
b_{21}=&&i2\pi\rho\left({\cosh_{\frac 12}}+\hat s\tau_3 {\sinh_{\frac 12}}\right)e^{i\pi x_4\tau_3}\nonumber\\
b_{12}=&&\left(-\cos(\pi x_4)(\cosh_{\frac 12}\overline{\sinh _{\frac 12}}\hat r +\cosh_{\frac 12}\overline{\sinh _{\frac 12}}\hat rs)\right.\nonumber\\&&\left.+ i\sin(\pi x_4) (\cosh_{\frac 12}\overline{\cosh _{\frac 12}} +\hat s\hat r \sinh_{\frac 12}\overline{\sinh _{\frac 12}})\right)
\nonumber\\
b_{22}=&&\left(-\cos(\pi x_4)(\cosh_{\frac 12}\overline{\sinh _{\frac 12}}\hat r +\cosh_{\frac 12}\overline{\sinh _{\frac 12}}\hat rs)\right.\nonumber\\&&\left.+ i\sin(\pi x_4) (\cosh_{\frac 12}\overline{\cosh _{\frac 12}} +\hat r\hat s \sinh_{\frac 12}\overline{\sinh _{\frac 12}})\right)
\nonumber\\\eea
and

\be
\psi\equiv -\cos(2\pi x_4)+ \cosh\overline{\cosh} +\frac {{\vec s}\cdot {\vec r}}{sr}
\sinh\overline{\sinh}\nonumber\\
\ee
with the short notation

\bea
\sinh_{\frac 12}=&&\sinh(\omega_0 \nu s)\nonumber\\
\cosh_{\frac 12}=&&\cosh(\omega_0 \nu s)\nonumber\\
\overline{\sinh_{\frac 12}}=&&\sinh(\omega_0 (1-\nu)r)\nonumber\\
\overline{\cosh_{\frac 12}}=&&\cosh(\omega_0 (1-\nu) r)
\eea

\section{Appendix D: Fock contribution}

In the main text, the mean-field analysis was presented using the so-called
Hartree approximation. Here we show how the Fock or exchange terms can be
included. We first that omitted crossed contractions in

\be
\left<\psi^T\epsilon \psi(x) \bar \psi^T\epsilon\bar \psi(y)\right>e^{i\phi_1^{\dagger}(x)+i\phi_1(y)}
\ee
can be retained by defining the $2\times 2$ propagator

\be
\label{PROPX}
\left<(\psi(x),\bar \psi(x))(\epsilon\psi^T(y),-\epsilon \bar \psi^T(y))^T\right>={\bf S}(x-y)
\ee
in terms of which the effective action $\mathbb S$ is a functional of (\ref{PROPX})

\bea
&&-{\mathbb S}[{\bf S},{\bf  b},\nu]={\rm Tr}\left({\bf S}_0^{-1}{\bf S}\right)-{\rm Tr} \ln {\bf S}\nonumber\\
&&+8\pi f_M \left(\frac{{\rm Tr}{\bf S}}{2}\right)^2\nu e^{\bf b}+8\pi f_L\bar \nu e^{-\bf b}\nonumber \\
&&-\frac{16\pi^2f_M^2}{T^3}\left(\frac{{\rm Tr}{\bf S}}{2}\right)^4\frac{1-e^{-V_0}}{\nu}e^{2\bf b}-\frac{16\pi^2f_L^2}{T^3}\frac{1-e^{-V_0}}{\bar \nu}e^{-2\bf b}\nonumber \\ &&+\frac{16\pi^2f_M^2}{T^3}e^{-V_0}e^{2\bf  b}\nonumber\\
&&\times\int_0^{\frac{x_0}{2\omega_0\nu}} d^3x\, {\rm Tr}({\bf S}^+_{12}(x){\bf S}^+_{21}(-x))
{\rm Tr}({\bf S}^-_{12}(x){\bf S}^-_{21}(-x))\nonumber\\
\eea
Here $S_{ij}$ are the pertinent entries in (\ref{PROPX}). The two gap equations are now extrema
of $\delta\mathbb S/\delta {\bf S}_{ij}=0$. If we were to approximate the term ${\rm Tr}({\bf SS})$ with free
propagators, then the gap equations simplify and we have for the dyonic part of the pressure

\bea
&&{\cal P_D}\rightarrow 8\pi f_M \nu\Sigma^2e^{2\bf b}+8\pi f_L \bar \nu e^{-\bf  b}\nonumber \\
&&-\frac{16 \pi^2f_M^2 }{T^3}\Sigma^4\frac{1-e^{-V_0}}{\nu}e^{2\bf b}-\frac{16 \pi^2 f_L^2}{T^3}\frac{1-e^{-V_0}}{\bar \nu}e^{-2\bf b}\nonumber \\&&+\frac{16\pi^2f_M^2}{T^3}e^{-V_0}e^{2\bf  b}\int_0^{\frac{x_0}{2\omega_0\nu}} d^3r\, {\rm Tr}( {\bf T}(r){\bf T}(-r))-4\Lambda \Sigma\nonumber\\
\eea

\section{Appendix E:  1-loop approximation}

An alternative to the mean-field analysis is based on the use of the 1-loop fermionic contribution
only. The 1-loop result is then used to compute the contractions induced by the second cumulant
contribution stemming from the core. The result for the constraint equation is

\bea
\label{SSS}
\Lambda({\bf b}, \nu)=2\pi\sqrt{f_Lf_M}(\nu e^{\bf b}+\bar \nu e^{-\bf b})
\eea
and the gap equation is

\bea
2\tilde \Sigma(\Lambda)=\pi\int \tilde p^2 d\tilde p\frac{\tilde \Lambda \mathbb F}{1+\tilde \Lambda^2\mathbb F}
\eea
To  1-loop the dressed fermionic propagator is

\bea
\label{SSSS}
S^{-1}=\tilde {\bf G}^{-1}+\Lambda({\bf b},\nu) \epsilon\left(\begin{array}{cccc}
0&1&0&0\\
1&0&0&0\\
0&0&0&-1\\
0&0&-1&0
\end{array}\right)
\eea
(\ref{SSS}-\ref{SSSS}) can be used  to reduce the contractions stemming from the second cumulant of the core,
as we detailed in section Vb. The result is an effective action solely dependent on ${\bf b}, \nu$, that is readily
analyzed in the weak coupling and strong screening limits. The results of this analysis will be
reported elsewhere.

\section{Appendix F: Holonomy potential}

For completeness, the instanton-dyon pressure with hopping fermions has to be supplemented with
the 1-loop perturbative contributions from the adjoint periodic gluons and anti-periodic fermions for
a finite holonomy $\nu$~\cite{OGILVIE}. The result for $N_f$ massless adjoint quarks is

\bea
\label{PRESSUREXX}
{\cal P}_{\rm 1loop} (N_f)=&&\frac{4T^3}{\pi^2}\sum_{n=1}^{\infty}\left(1-N_f(-1)^n\right)\frac{{\rm Tr}_AL^n}{n^4} \nonumber\\
{\cal P}_{\rm 1loop}(1)=&&\frac{16T^3}{\pi^2}\sum_{n=0}\frac{\cos (4n+2)\pi \nu }{(2n+1)^4}
\eea
with $L=e^{i2\pi \nu T_3}$. The first contribution is from the adjoint gluons
while the second contribution is from the anti-periodic adjoint fermions.
The perturbative minima of (\ref{PRESSUREXX}) at $\nu=0,1$
yields a finite Polyakov line or an asymmetric (non-confining) ground state.
Note that for $N_f=1$ periodic adjoint fermions $(-1)^n\rightarrow 1$ in (\ref{PRESSUREXX})
and the bosonic and fermionic contributions cancel out. This result is expected from
supersymmetry.

 \vfil

\end{document}